\documentclass[aps,prd,twocolumn,nofootinbib,showpacs,preprintnumbers]{revtex4-1}
\usepackage{natbib}
\usepackage{longtable}
\usepackage[pdftex]{graphicx}
\usepackage[usenames]{color}
\usepackage[dvips]{epsfig}

\usepackage{hyperref}
\usepackage{bbm}
\usepackage{marvosym}
\usepackage{verbatim}
\usepackage{psfrag}
\usepackage{slashed}

\usepackage{amsmath}
\usepackage{amssymb}
\usepackage{wasysym}
\usepackage{multirow}

\usepackage{diagbox}
\usepackage{siunitx} 
\usepackage{array} 
\usepackage{amssymb}	
\usepackage{booktabs,amsmath}
\usepackage{bbold}
\usepackage[table,x11names]{xcolor}
\definecolor{maroon}{cmyk}{0,0.87,0.68,0.32}
\usepackage{amsfonts}

\definecolor{boxcolor}{HTML}{ffe6a1}

\newcommand{\Ord}[1]{\mathcal{O}\left(#1\right)}

\newcommand{\eref}[1]{Eq.~(\ref{#1})}
\newcommand{\fref}[1]{Fig.~\ref{#1}}
\newcommand{\sref}[1]{Section~\ref{#1}}
\newcommand{\tref}[1]{Table~\ref{#1}}

\def \Nt {{N_{\tau}}}
\def \Nc {{N_{c}}}
\def \Nf {{N_{f}}}

\newcommand{\expval}[1]{\langle #1 \rangle}

\newcommand{\Ud}{U^\dagger}
\newcommand{\M}{\mathcal{M}}
\newcommand{\Md}{\mathcal{M}^\dagger}

\newcommand{\calZ}{\mathcal{Z}}
\newcommand{\calD}{\mathcal{D}}

\newcommand{\lr}[1]{\left( #1 \right)}

\newcommand{\nn} {\nonumber\\}
\newcommand{\beqn} {\begin{equation}}
\newcommand{\eqn} {\end{equation}}

\def \beq{\begin{equation}}
\def \eeq{\end{equation}}
\def \bea{\begin{eqnarray}}
\def \eea{\end{eqnarray}}

\def \Tr {{\rm Tr}}
\def \tr {{\rm tr}}

\def \bet0{\beta_0}
\def \bet1{\beta_1}
\def \simgt{\,\rlap{\lower 7.5 pt\hbox{$\mathchar \sim$}}\raise 3 pt \hbox{$>$}\,}
\def \simlt{\,\rlap{\lower 7.5 pt\hbox{$\mathchar \sim$}}\raise 3 pt \hbox{$<$}\,}

\def\lsim{\raise0.3ex\hbox{$<$\kern-0.75em\raise-1.1ex\hbox{$\sim$}}}
\def\gsim{\raise0.3ex\hbox{$>$\kern-0.75em\raise-1.1ex\hbox{$\sim$}}}

\newcommand{\SU}{{\rm SU}}
\newcommand{\U}{{\rm U}}

\newcommand{\hmu}{{\hat{\mu}}}
\newcommand{\hnu}{{\hat{\nu}}}

\DeclareMathAlphabet{\mathpzc}{OT1}{pzc}{m} {it}

\begin{document}
    
\title{Nuclear Liquid-Gas Transition in the Strong Coupling Regime of Lattice QCD}
\author{J.~Kim$^{\rm a}$}
\email{j.kim@fz-juelich.de}
\author{P.~Pattanaik$^{\rm b}$}
\email{pratiteep@physik.uni-bielefeld.de}
\author{W.~Unger$^{\rm b}$}
\email{wunger@physik.uni-bielefeld.de}
\affiliation{$^{\rm a}$ Institute for Advanced Simulation (IAS-4), Forschungszentrum J\"ulich, Wilhelm-Johnen-Stra\ss e, 52428 J\"ulich, Germany} 
\affiliation{$^{\rm b}$ Fakult\"at f\"ur Physik, Bielefeld University, D-33615 Bielefeld, Germany}

\begin{abstract}
The nuclear liquid-gas transition from a gas of hadrons to a nuclear phase cannot be determined numerically from conventional lattice QCD due to the severe sign problem at large values of the baryon chemical potential. In the strong coupling regime of lattice QCD with staggered quarks, the dual formulation is suitable to address the nuclear liquid gas transition. We determine this first order transition at low temperatures and as a function of the quark mass and the inverse gauge coupling $\beta$.
We also determine the baryon mass and discuss the nuclear interactions as a function of the quark mass, and compare to mean field results.
\end{abstract}

\maketitle
\tableofcontents

\section{Introduction}

It is known from experiments \cite{Natowitz2002} that at low temperatures, there is a phase transition between dilute hadron gas and dense nuclear matter as the baryon chemical potential increases. This transition is of first order and terminates at about $T_c=16$ MeV in a critical end point.
The value of the chemical potential $\mu_B^{1st}$ at zero temperature is given roughly by the baryon mass $m_B$, where the difference of $\mu_B^{1st}-m_B$ is due to nuclear interactions. For a review on nuclear interactions see \cite{Epelbaum2008}.

As the nuclear force between baryons to form nuclear matter is due to the residual strong interactions between quarks and gluons, it should be accurately described by QCD. We choose to study the nuclear transition and nuclear interaction via lattice QCD \cite{Ishii2007}, with its Lagrangian being a function of the quark mass and the inverse gauge coupling. In order to understand the nature of the transition, it is helpful to study its dependence on these parameters. 

However, at finite baryon density, lattice QCD has the infamous sign problem which does not allow us to perform direct Monte Carlo simulations on the lattice. Various methods have been proposed to overcome the numerical sign problem, but they are either limited to  $\mu_B/T\lesssim 3$ \cite{Fodor2001, deForcrand2003,DElia2002,Allton2005} or can not yet address full QCD in 3+1 dimensions in the whole  $\mu_B - T$ plane 
\cite{Cristoforetti2012, Sexty2013}, in particular the nuclear transition is out of reach. 

An alternative method is to study lattice QCD via the strong coupling expansion. There are two established effective theories for lattice QCD based on this: (1) the 3-dim.~effective theory for Wilson fermions in terms of Polyakov loops, arising from a joint strong coupling and hopping parameter expansion \cite{Langelage2014}, (2) the dual representation for staggered fermions in 3+1 dimensions, with dual degrees of freedom describing mesons and baryons.
Both effective theories have their limitations: (1) is limited to rather heavy quarks (but is valid for large values of $\beta$) whereas (2) is limited to the strong coupling regime $\beta \lesssim 1$ (but is valid for any quark mass). 
We study lattice QCD in the dual formulation, both at infinite bare gauge coupling, $\beta = 0$, and at leading order of the strong coupling expansion in the regime $\beta<1$, which is far from the continuum limit.
But since strong coupling lattice QCD shares important features with QCD, such as confinement, and chiral symmetry breaking and its restoration at the chiral transition temperature, and a nuclear liquid gas transition, we may get insights into the mechanisms, in particular as the dual variables give more information in terms of its world lines, as compared to the usual fermion determinant that depends on the gauge variables.

To establish a region of overlap of both effective theories, we have chosen to perform the Monte Carlo simulations in the dual formulation extending to rather large quark masses. 

This paper is organized as follows: in the first part we explain the dual formulation in the strong coupling regime, in the second part we provide analytic results based on exact enumeration and mean field theory, in the third part we explain the setup of our Monte Carlo simulations and present result on the $m_q$- and $\beta$-dependence of the nuclear transition. 
Since the strong coupling regime does not have a well defined lattice spacing, we also determine the baryon mass $am_B$ to set the parameters of the grand-canonical partition function, $aT$ and $a\mu_B$, in units of $am_B$.
We conclude by discussing the resulting nuclear interactions, and compare our findings with other results.

\subsection{Staggered action of strong coupling QCD and its dual representation}

In the strong coupling regime, the gauge integration is performed first, followed by the Grassmann integration to obtain a dual formulation. This was pioneered for the strong coupling limit in \cite{Rossi1984} and has been extended by one of us to include gauge corrections \cite{deForcrand2014,Gagliardi2019}. The sign problem is mild in the strong coupling limit and still under control for $\beta<1$, where we can apply sign reweighting. The dual degrees of freedom are color-singlet mesons and baryons, which are point-like in the strong coupling limit, and become extended about a lattice spacing by incorporating leading order gauge corrections.

The partition function of lattice QCD is given by 
\beq
\calZ=\int \calD \U \calD\bar{\chi} \calD \chi e^{-S_G[\U]-S_F[\bar{\chi}, \chi, \U]}
\eeq
where $\calD \U$ is the Haar measure, $U\in\SU(3)$ are the gauge fields on the lattice links ($x,\hat{\mu}$) and $\{\bar{\chi_x},\chi_x\}$ are the unrooted staggered fermions at the lattice sites $x$. The gauge action $S_G[\U]$ is given by the Wilson plaquette action 
\begin{align}
S_G[\U]&=-\frac{\beta}{2 \Nc} \sum_p \Tr[\U_p]+\Tr[\U_p^{\dagger}]
\label{WilsonGaugeAction}
\end{align}
and the staggered fermion action $S_F[\bar{\chi}, \chi, \U]$ is:
\begin{widetext}
\begin{align}
S_F[\bar{\chi}, \chi, \U]&=\sum_{x, \hmu} \eta_\hmu(x)\left(e^{+a\mu_q \delta_{\hmu, \hat{0}}} \bar{\chi}_x \U_{x, \hmu} \chi_{x+\hmu}-e^{-a\mu_q \delta_{\hmu, \hat{0}}} \bar{\chi}_{x+\hmu} \U_{x, \hmu}^{\dagger} \chi_x\right) + 2am_q\bar{\chi}_x \chi_x,
\end{align}
\end{widetext}
where the gauge action depends on the inverse gauge coupling $\beta=\frac{2\Nc}{g^2}$ and the fermion action depends on the quark chemical potential $a\mu_q$ which favors quarks in the positive temporal direction, and the bare quark mass $am_q$.\\

First we consider the strong coupling limit where the inverse gauge coupling $\beta$=0 and hence the gauge action $S_G[\U]$ drops out from the partition function in this limit. 
The gauge integration is over terms depending only on the individual links ($x,\hmu$) so the partition function factorizes into a product of one-link integrals and we can write it as:
\begin{widetext}
\begin{align}
\calZ &= \int \prod_{x}\left(\mathrm{~d} \chi_{x} \mathrm{~d}_{\chi_{x}} \mathrm{e}^{2 a m_{q} \bar{\chi}_{x} \chi_{x}} \prod_{\hat{\mu}} z(x, \hat{\mu})\right),& z(x,\hmu)&=\mathrm{d} \U_{\hat{\mu},x} \mathrm{e}^{\eta_{\hat{\mu}}(x)\left(\bar{\chi}_{x} \U_{\hat{\mu},x} \chi_{x+\hat{\mu}}-\bar{\chi}_{x+\hat{\mu}} \U_{\hat{\mu},x}^{\dagger} \chi_{x}\right)},
\label{LinkFactorization}
\end{align}
\end{widetext}
with $z(x,\hat{\mu})$ the one-link gauge integral that can be evaluated from invariant integration, as  discussed in \cite{Creutz1978,Rossi1984}, where we write the one-link integral in terms of new hadronic variables:
\begin{align}
M(x)&=\bar{\chi}_{x} \chi_{x},& B(x)&=\frac{1}{N!} \varepsilon_{i_{1} \ldots i_{\Nc}} \chi_{x, i_{1}} \cdots \chi_{x, i_{\Nc}}
\end{align}
Only terms of the form $(M(x)M(y))^{k_{x,\hmu}}$ (with $k_{x,\hmu}$ called dimers which count the number of meson hoppings) and $\bar{B}(y)B(x)$ and $\bar{B}(x)B(y)$ (called baryon links) are present in the solution of the one-link integral. The sites $x$ and $y=x+\hmu$ are adjacent lattice sites. It remains to perform the Grassmann integral of the fermion fields $\bar{\chi}$, $\chi$. This requires to expand the exponential containing the quark mass in \eref{LinkFactorization} (left), which results in the terms $(2am_qM(x))^{n_x}$ (with $n_x$ called monomers). To obtain non-vanishing results, at every site, the $2\Nc$ Grassman variables $\chi_{x,i}$ and $\bar{\chi}_{x,i}$ have to appear exactly once, resulting in the Grassmann constraint (GC): 
\begin{align}
n_x+\sum_{\pm \hat{\mu}} \left( k_{x,\hat{\mu}} + \frac{\Nc}{2} |\ell_{x,\hat{\mu}}| \right ) = \Nc,
\label{GrassmannC}
\end{align}
where $n_x$ is the number of monomers, $k_{x,\hat{\mu}}$ is the number of dimers and the baryons form self-avoiding loops $\ell_{x,\hat{\mu}}$, which due to the constraint cannot coexist with monomers or dimers.

With this, we obtain an exact  rewriting of the partition function \eref{LinkFactorization} for $N_c=3$, in terms of integer-valued dual degrees of freedom $\{n,k,\ell\}$:
\begin{widetext}
\begin{align}
\calZ=\sum_{\{k, n, \ell\}}^{\rm GC} \prod_{b=(x, \hat{\mu})} \frac{\left(3-k_{b}\right) !}{3 ! k_{b} !} \gamma^{2 k_{b} \delta_{0, \hat{\mu}}} \prod_{x} \frac{3 !}{n_{x} !}\left(2 a m_{q}\right)^{n_{x}} \prod_{\ell} w(\ell)
\label{DualPF}
\end{align}
\end{widetext}
where the sum over valid configurations has to respect the constraint (GC).
The first term in the partition function is the contribution from dimers and the second term is the contribution from monomers. The weight factor $w(\ell)$ for each baryon loop $\ell$ depends on the baryon chemical potential $\mu_B=3\mu_q$ and induces a sign factor $\sigma(\ell)$ which depends on the geometry of $\ell$:
\begin{align}
w(\ell)=\frac{1}{\prod_{x \in \ell} 3 !}\sigma(\ell) \gamma^{3 N_{\hat{0}}} \exp \left(\omega_{\ell} N_{t} a_{t} \mu_B\right).
\end{align}
Here, $\omega_\ell$ is the winding number of the loop $\ell$.
The total sign factor $\prod_\ell \sigma(\ell)\in \{\pm 1\}$ is explicitly calculated for every configuration.
We apply sign reweighting as the dual formulation has a mild sign problem: baryons are non-relativistic and usually have loop geometries that have a positive signs. The dual partition function of the strong coupling limit is simulated with the worm algorithm (see \sref{MonteCarlo}) and the sign problem is essentially solved in this limit.\\

\subsection{Extension to finite $\beta$}

The leading order gauge corrections $\mathcal{O}(\beta)$ to the strong coupling limit are obtained by expanding the Wilson gauge action Eq.~(\ref{WilsonGaugeAction})
before integrating out the gauge links. A formal expression is obtained by changing the order of integration (first gauge links, then Grassmann-valued fermions) 
within the QCD partition function: 
 \begin{align}
  Z_{QCD} &= \int d\chi d\bar{\chi} DU e^{-S_G[U]-S_F[U]}\nn
  &= \int d\chi d\bar{\chi} Z_F  \expval{e^{-S_G[U]}}_{Z_F},& 
  Z_{F} &= \int DU e^{-S_F[U]}.
 \end{align}
 With this the $\Ord{\beta}$ partition function is
 \begin{align}
  Z^{(1)} &= \int d\chi d\bar{\chi} Z_F \expval{-S_G[U]}_{Z_F},\\
  \expval{-S_G[U]}_{Z_F}&=\frac{\beta}{2\Nc}\frac{\int DU \sum_P \lr{\tr[ U_P+U_P^\dagger]}e^{-{S_F[U]}}}{Z_F}.
 \end{align}
 The challenge in computing $Z^{(1)}$ is to address the $\SU(\Nc)$ integrals that receive contributions from the elementary plaquette $U_P$. 
 Link integration no longer factorizes, however the $\tr[U_P]$ can be decomposed before integration:
 \begin{align}
  \int DU \tr[ U_P]e^{-{S_F[U]}}&= J_{ab}J_{bc}J_{cd}J_{da},\nn
  J_{ij}(\M,\Md)&=\int DU \,e^{\tr[U\Md+\M\Ud]}\,U_{ij}\
 \end{align}
 Integrals of the type $J_{ij}$ with two open color indices - as compared to link integration at strong coupling - have been derived from
 generating functions
 \begin{align}
 Z^{a,b}[K,J] &= \int_G DU \tr[UK ]^a \tr[U^\dagger J ]^b
 \end{align}
 for either $J=0$ \cite{Creutz1978} or for $G=\U(\Nc)$ \cite{Eriksson1981,Azakov1988}. 
 The $\SU(3)$ result was discussed in \cite{deForcrand2014}, 
 in terms of the dual variables, neglecting rotation and reflection symmetries, there are 19 distinct diagrams to be considered. The resulting partition function, valid to $\mathcal{O}(\beta)$, is
\begin{align}
Z(\beta) &=\sum_{\{n,k,\ell,q_P\}}\prod_x \hat{w}_x \prod_{b} \hat{w}_b \prod_\ell \hat{w}_\ell \prod_P \hat{w}_P, \nn
\hat{w}_x & = w_x v_x,\qquad \hat{w}_b = w_b k_b^{q_b}, \qquad \hat{w}_\ell = w_\ell \prod_\ell { w_{B_i}(\ell)},\nn
\hat{w}_P&=\left(\frac{\beta}{2N}\right)^{|q_P|},
\label{ParFuncOB}
\end{align}
with $q_P\in\{0,\pm 1\}$, and
the site weights $w_x \mapsto \hat{w}_x$, bond weights $w_b \mapsto \hat{w}_b$ and baryon loop weights $w_\ell \mapsto \hat{w}_\ell$ receive modifications 
compared to the strong coupling limit \eref{DualPF} for sites and bonds adjacent to an excited plaquette $q_P=1$.
The weights are given in \cite{deForcrand2014}, and are re-derived for any gauge group in \cite{Gagliardi2019}.
The configurations $\{n,k,\ell,q_p\}$ must satisfy at each site $x$ the constraint
inherited from Grassmann integration:
\begin{equation}
n_x+\sum_{\hnu=\pm\hat{0},\ldots, \pm \hat{d}}\lr{k_{\hnu}(x) + \frac{N_c}{2} |\ell_\hnu(x)|} = N_c+q_x,
\label{GrassmannOB}
\end{equation}
which is the modified version of \eref{GrassmannC} with $q_x=1$ if located at the corner of an excited plaquette $q_p\neq 0$, otherwise $q_x=0$.

\begin{figure}[h!]
    \centering
\includegraphics[width=\columnwidth]{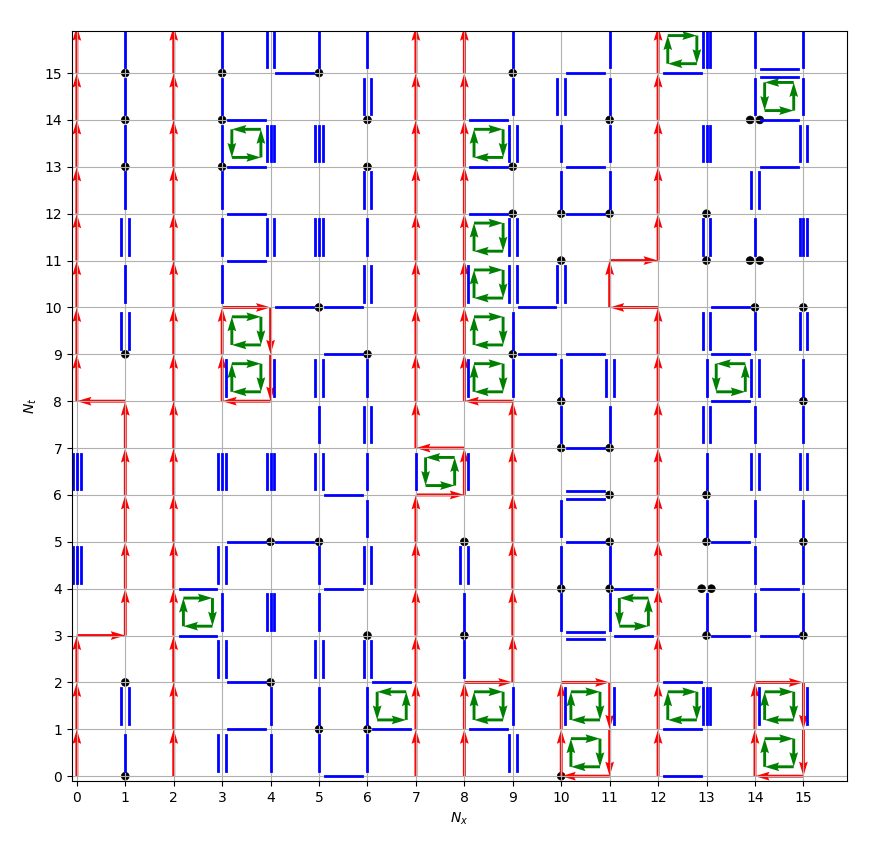}
    \caption{Typical 2-dimension configuration at $\beta=1.0$, at non-zero quark mass, temperature, chemical potential.
    The black dots are monomers, the blue lines are dimers, the red arrows are baryon loop segments (or triplets $g_b+f_b=\pm3$ if adjacent to a non-trivial plaquette), and the green squares are plaquette occupations $\pm 1$.
    The actual configurations are 3+1-dimensional.
    }
    \label{Config}
\end{figure}

A more general expression that we obtained via group theory and is valid to higher orders of the strong coupling expansion is discussed in terms of tensor networks \cite{Gagliardi2019}. 
A typical 2-dimensional configuration that arises at $\beta=1$ in the Monte Carlo simulations is given in \fref{Config}.
Note that if a baryon loop enters a non-trivial plaquette, one quark is separated from the two other quarks, resulting in the baryon being extended object, rather being point-like in the strong coupling limit. 

The $\mathcal{O}(\beta)$ partition function has been used in the chiral limit \cite{deForcrand2014} to study the full $\mu_B-T$ plane via reweighting from the strong coupling ensemble. Whereas the second order chiral transition for small values of the $a\mu_B$ decreased up to the tri-critical point, the first order nuclear transition was invariant:  $a\mu_B^{1st}\simeq 1.78(1)$ at zero temperature has no $\beta$-dependence. 
For the ratio $T(\mu_B=0)/\mu_B^{1st}(T\simeq 0)$ we found the values $0.787$ for $\beta=0$ and  $0.529$ $\beta=1$, which should be compared to $T_c/\simeq 0.165$ for full QCD \cite{Bazavov2011}. 

However, since reweighting cannot be fully trusted across a first order boundary, direct simulations at non-zero $\beta$ are necessary. The Monte Carlo technique to update plaquette variables is discussed in \sref{MonteCarlo}.

\section{Analytic results}
In this section, we provide analytic results from exact enumeration for small volumes, and mean field results based on the 1/d expansion, valid in the thermodynamic limit. The main purpose is to compare our Monte Carlo results to these analytic predictions. 

\subsection{Exact enumeration}
To establish that our Monte Carlo simulations indeed sample the partition functions \eref{DualPF} and \eref{ParFuncOB}, we have obtained analytic results on a $2^4$ volume at strong coupling, and at finite beta in two dimensions on a $4 \times 4$ volume, comparing $\Ord{\beta}$ and $\Ord{\beta^2}$ truncations.

Our strategy to obtain an exact enumeration of the partition function $Z$ is to enumerate plaquette configurations first, then fixing the fermion fluxes which together with the gauge fluxes that are induced by the plaquettes form a singlet, a triplet or anti-triplet, i.e.~on a given bond $b$, $g_b+f_b \in \{-3,0,3\}$, and last we perform the monomer-dimer enumeration on the available sites not saturated by fermions yet by a depth-first algorithm \cite{Krauth2006}. At strong coupling, with no plaquettes, $g_b=0$ and $f_b$ are baryonic fluxes.

\begin{figure}[!ht]
    \centering
    \includegraphics[width=\columnwidth,page=1]{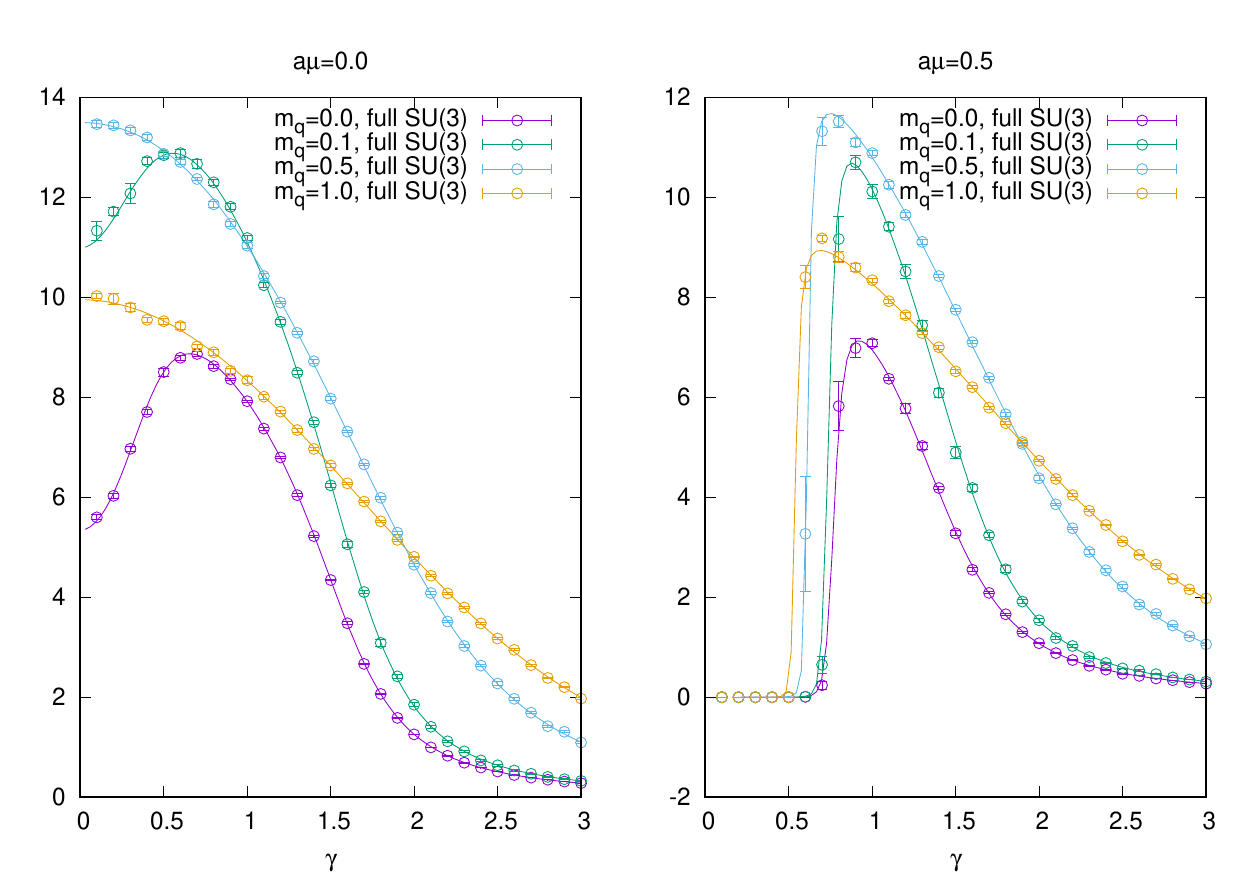} \caption{Chiral susceptibility on a $2^4$ volume for various quark masses, as a function of the bare anisotropy $\gamma$ (with $aT=\gamma^2/2$), analytic results from enumeration compared to numerical data from simulations via the worm algorithm.}
    \label{Enumeration1}
\end{figure}

\begin{figure*}[ht!]
    \centering
\includegraphics[width=\textwidth]{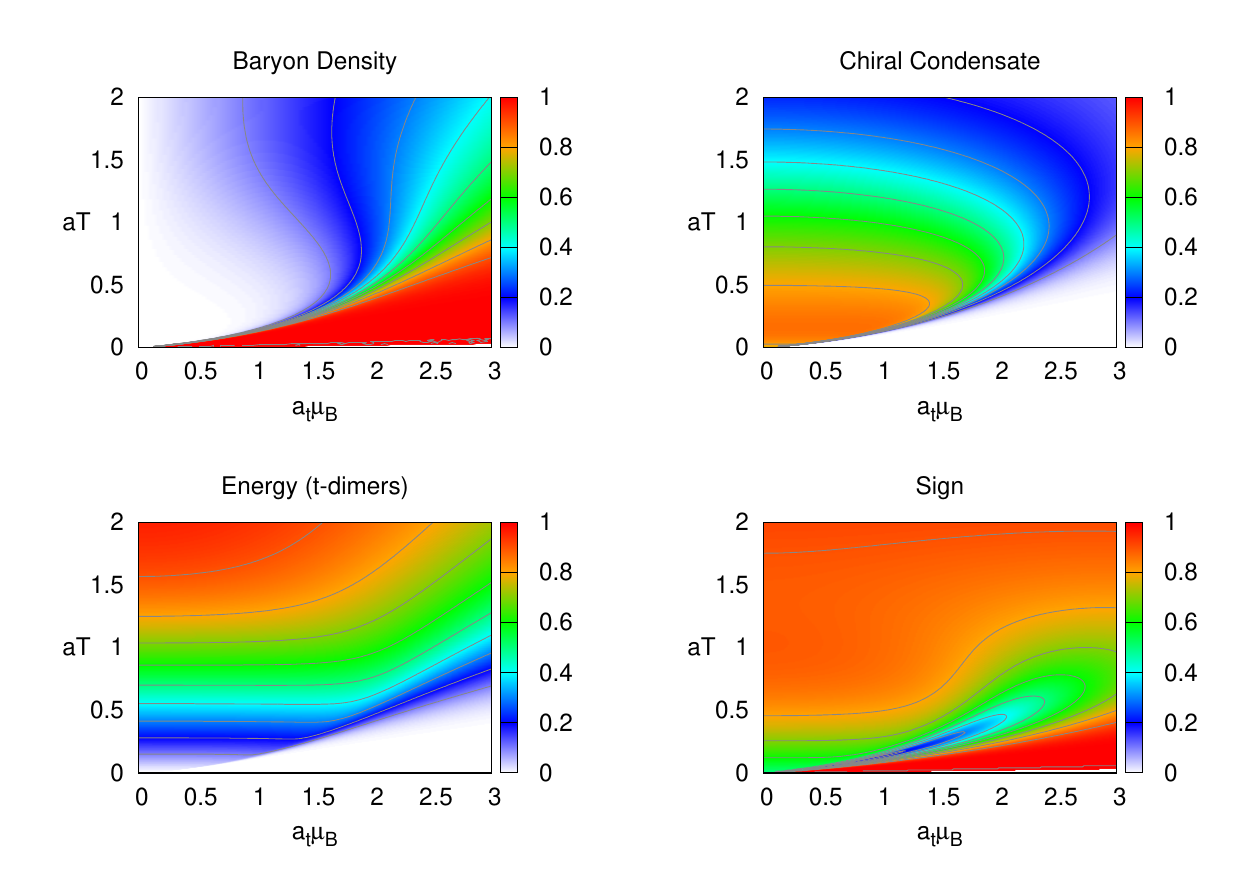}
    \caption{Various observables in the $\mu_B$-$T$ plane on a $2^4$ volume at $am_q=0.1$. The back-bending of the first order transition at temperatures below $aT=0.5$ in all observables is an artifact of the small volume, and vanishes in the thermodynamic limit. The temperature $aT=1/2$ corresponds to the isotropic lattice here. 
    }
    \label{Enumeration2}
\end{figure*}

All observables that can be written in terms of derivatives of $\log(z)$, such as the baryon density, the chiral condensate, the energy density, and also the average sign, are shown in \fref{Enumeration2} in the full $\mu_B - T$ plane. A detailed comparison of Monte Carlo data with exact enumeration is shown in \fref{Enumeration1}. 

\subsection{Expectations from mean field theory}
\label{MF}

Another analytical method to study strong coupling lattice QCD is the mean field approach, where the partition function is expanded in $\frac{1}{d}$ ($d$ is the spatial dimension) and then a Hubbard-Stratonovich transformation performed \cite{Nishida2003}. After this procedure, the free energy is a function of temperature $T$, the chiral condensate $\sigma$ and chemical potential $\mu_B$:
\begin{widetext}
\begin{align}
F_{\text {eff }}\left[\sigma, T,m, \mu_{\mathrm{B}}\right]&=\frac{N_{\mathrm{c}} d}{4}\sigma^{2}-T \log \left\{2 \cosh \left[\mu_{\mathrm{B}} / T\right]+\frac{\sinh \left[\left(N_{\mathrm{c}}+1\right) E / T\right]}{\sinh [E / T]}\right\},\nn
E[m]&=\text{arcsinh}\left[\sqrt{m^2+\left(\frac{d\sigma}{2}\right)^2+m d\sigma}\right],
\end{align}
\end{widetext}
here $E[m]$ is one-dimensional quark excitation energy which is a function of the quark mass $m=am_q$. For  $\Nc=3$ and $d=3$ we determined the minimum of the free energy with respect to the chiral condensate. This gives us the equilibrium chiral condensate as a function of $(T,m,\mu_B)$.

\begin{figure}[h!]
\centering
\includegraphics[width=\columnwidth]{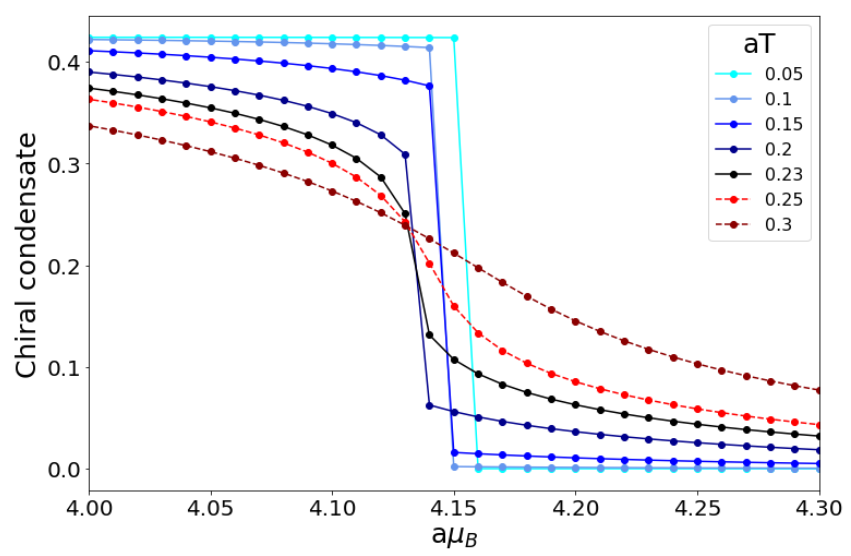}
\includegraphics[width=\columnwidth]{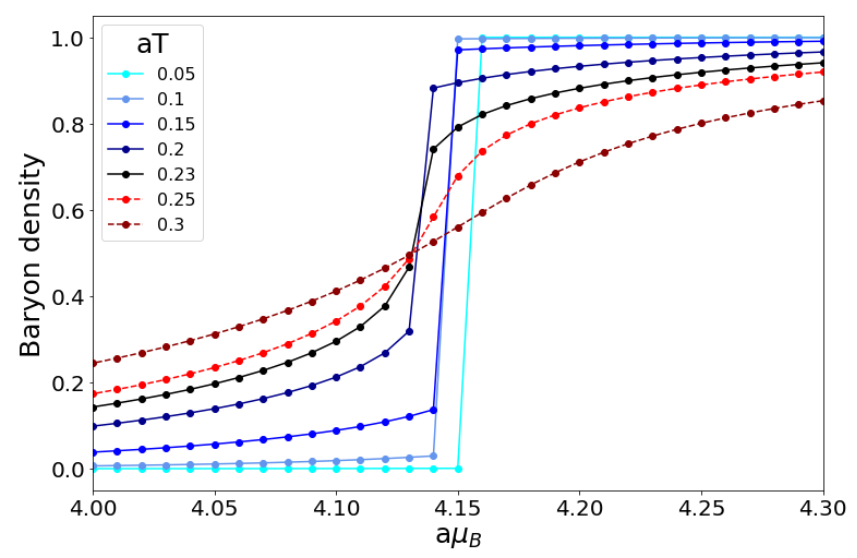}
\caption{The chiral condensate (\emph{left}) and the baryon density (\emph{right}) for quark mass $m=1.5$ as a function of the chemical potential and for various temperatures.}
\label{chiral_condensate}
\end{figure}

The chiral condensate and the baryon density as a function of the baryon chemical potential in lattice units $a\mu_B$ and for various temperatures at quark mass $m=1.5$ is shown in \fref{chiral_condensate}.  We have determined the critical temperature to be $aT_c=0.23(1)$, which is characterized by an infinite slope of the chiral condensate. For lower temperatures, there is a clear discontinuity of the chiral condensate, separating the low density phase from the high density phase. For temperatures above and in the vicinity of $aT_c$ the chiral condensate and baryon density has no discontinuity but rapidly changes, corresponding to a crossover transition.

\begin{figure}[h!]
  \centering
\includegraphics[width=\columnwidth]{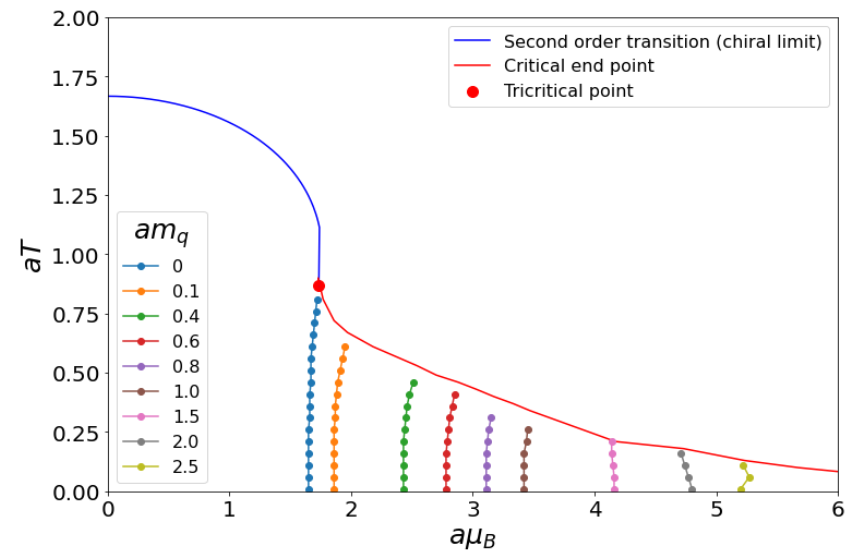}
\caption{Phase diagram with mean field approach. The second order chiral transition is shown in blue (solid line). The phase transition for different quark masses are the plotted with dotted lines. The critical end point for different quark masses is plotted in red (solid line).}
\label{phase_diagram}
\end{figure}

With this method, the phase diagram is plotted for different quark masses in \fref{phase_diagram}. The second order phase transition in the chiral limit is plotted in solid blue line, the dotted lines show the first order phase transition for different quark masses and the solid red line indicates the critical end point for the different quark masses.

Mean field theory also gives an expression for the pion mass $am_\pi$ and the baryon mass $am_B$ \cite{KlubergStern1983b}:
\begin{widetext}
\begin{align}
 (am_\pi)^2 &= 2\sqrt{2d+2}\,m\stackrel{d=3}{=} 2\sqrt{2d+2}\,m,\\
\sinh \, (am_B) &= \frac{(2d+2)^{\Nc/2}}{2} \left(\frac{m}{\sqrt{2d+2}}+\sqrt{\frac{m^2}{2d+2}+1}\right)^{\Nc}
\stackrel{d=3,\Nc=3}{=} 4\sqrt{8} \left(\frac{m}{\sqrt 8}+\sqrt{\frac{m^2}{8}+1}\right)^{3}.
\end{align}
\end{widetext}
The mean field baryon mass for $\Nc=3$,\linebreak  $d=3$ is also plotted in red in \fref{comparison}. Whereas the baryon mass is around $\Nc$  in the chiral limit \linebreak($am_B\simeq 3.12$ for  $\Nc=3$), it approximately doubles at $m=3.5$ ($am_B\simeq 6.28$) which corresponds to the pion mass $am_\pi=4.45$, i.e. 
\linebreak$m_\pi/m_B=0.708$. 
Hence, at around bare mass $m=3.5$, the valence quark mass of the baryon corresponds roughly to $1/3$ of the chiral limit value of the baryon mass.

\section{Numerical Investigation}

\subsection{Monte Carlo Simulations}
\label{MonteCarlo}

The first Monte Carlo simulations that could extend in the $\mu_B-T$ plane was the MDP algorithm \cite{Karsch1989}, but it required the introduction of the worm algorithm \cite{Prokofev2001} to make substantial progress. First studies of the worm algorithm applied to the strong coupling limit QCD (with gauge group $\U(3)$) are \cite{Adams2003}, and \cite{Fromm2010,Fromm2009} for gauge group $\SU(3)$.
Monte Carlo simulations to extend the worm to incorporate leading order corrections were first proposed in \cite{deForcrand2014}.
We will shortly review the setup of or Monte Carlo strategy for the nuclear transition, with an emphasis on the challenges to address large quark masses.

\subsubsection{Strong Coupling}

In order to sample the dual variables in \eref{DualPF}, 
the worm algorithm performs much better concerning critical slowing down as other local algorithms. 
Although the worm (in spin models) is based on a high-temperature expansion (in our context this corresponds to the expansion in baryon and meson hoppings), it is valid for any temperatures (in our context: for all quark masses). 
The main idea of the worm is to sample an enlarged configuration space by introducing two sources known as worm tail $x_T$ and head $x_H$, and by proposing local updates to either move both head and tail, or shift the worm head $x_H$ until it recombines again with the tail, 
$x_T=x_H$. During worm evolution, the 2-point monomer correlation function is measured, and after the worm update has completed, a global update has occurred.

We use two types of worm evolutions, one for the mesonic sector (not touching baryonic sites), and one worm to modify, construct or deconstruct baryonic loops as explained in detail in \cite{Fromm2010}. This can be readily used to perform simulations at nonzero temperature and baryon density.  

Simulations in the chiral limit are particularly cheap with the worm algorithm. 
Finite quark masses also requires to include a monomer-dimer update that change the monomer number. This update is sufficient for small quark masses, but not for large quark masses $am_q>1.0$, in particular at low temperatures and densities around $a\mu_B,c$: 
the reason is that the quark mass favors monomers, the chemical potential favors baryons, which makes it difficult to pinpoint the first order transitions between the hadron gas and the nuclear phase, due to large autocorrelation times for baryonic observables.

To overcome this limitation, we propose an additional update, a 'static update', that is based on the 1-dim QCD partition function \cite{Bilic1988} and applies to all spatial sites with no spatial dimers or baryons attached at any $\tau \in [0,1,\ldots \Nt-1]$. This additional update drastically reduced the autocorrelation time by effectively mixing monomers and static baryons.

The particular quark masses, volumes, chemical potentials and the statistics in terms of the number of worm updates is given in \tref{SCTable}.

\subsubsection{Finite $\beta$}

Whereas the reweighting result could not answer the question about the $\beta$-dependence of the nuclear transition, 
direct simulations at finite $\beta$ could in principle resolve this issue. This required to implement a plaquette update based on the 
plaquette and anti-plaquette occupation numbers $n_p$, $\bar{n}_p$, which is essentially a Metropolis-Hasting algorithm.
We have restricted here to $n_p-\bar{n_p}=q_p\in\{0,\pm1\}$ as given in \eref{ParFuncOB} to have a manageable update strategy:
The mesonic and baryonic worm algorithms
need to be mixed with plaquette updates sufficiently to have an ergodic algorithm valid for sufficiently large $\beta$. In practice, after the worm closes,
a plaquette update is proposed on random plaquette coordinates $p$.
Typical configurations in terms of monomers, dimers, fermion world-lines and plaquette excitations as shown in \fref{Config}  are based on \eref{ParFuncOB}.
In practice, such simulations are limited by the sign problem which already becomes severe for $\beta>1$, see \sref{SignProblemOBeta}.

The particular quark masses, values of $\beta$, volumes, chemical potentials and the statistics in terms of the number of worm updates is given in \sref{OBTable}.

\subsection{Residual sign problem}

Although it is possible to resum the sign problem at strong coupling with a resummation of baryon and pion world lines, this is not possible when including gauge corrections. In order to compare both sign problems, we kept the original dual formulation to monitor the severity of the sign problem. This is done via the relation
\begin{align}
\expval{\sigma} &= e^{-V/T\Delta f},& \Delta_f=f-f_{||}
\end{align}
between the average sign $\expval{\sigma}$ and the difference of the free energy density $\Delta f$ between the full ensemble $f$ and of the sign-quenched ensemble $f_{||}$.

\subsubsection{Strong Coupling}
\begin{figure}[h!]
 \centering
\includegraphics[width=\columnwidth]{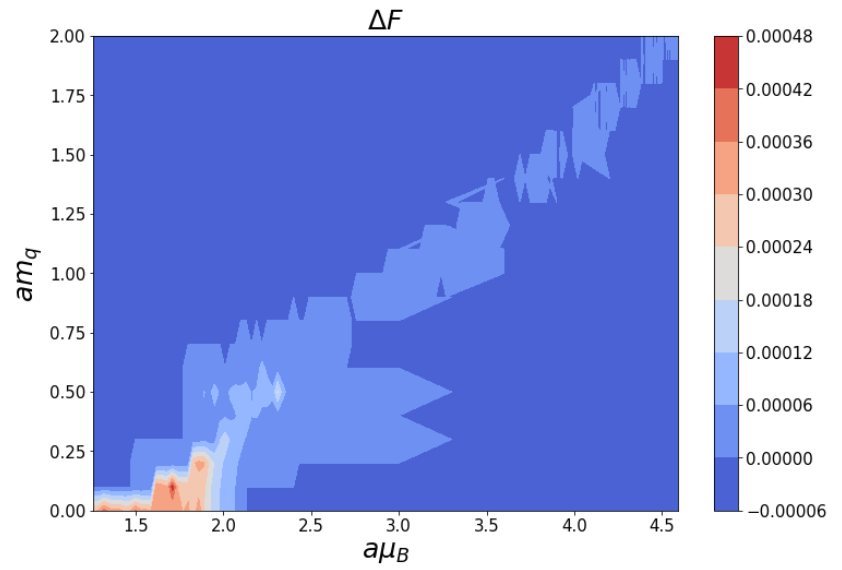}
\caption{$\Delta F$ at strong coupling as a function of chemical potential and quark mass on a $6^3\times 8$. The sign problem becomes milder as the quark mass increases.}
\label{SC sign}
\end{figure}

Without any further resummation, there is a mild sign problem in the dual formulation of lattice QCD in the strong coupling limit. When the average sign $\expval{\sigma}$ is not too small (close to zero), it implies that most of the configurations have a positive weight thus allowing us to perform sign reweighting strategies. In \fref{SC sign}, $\Delta f$ is plotted as a function of the baryon chemical potential and the quark masses. It is seen that $\Delta f$ is close to zero for most cases except near the critical chemical potential and for small quark masses, but never exceeds $5\times 10^{-4}$. Hence sign reweighting can be performed in the full parameter space. The result that the sign problem becomes even milder when increasing the mass is related to the fact that larger critical chemical potentials result in a larger fraction of static baryons (spatial baryon hoppings become rare).

\subsubsection{Finite $\beta$}
\label{SignProblemOBeta}

\begin{figure}[h!]
  \centering
\includegraphics[width=\columnwidth]{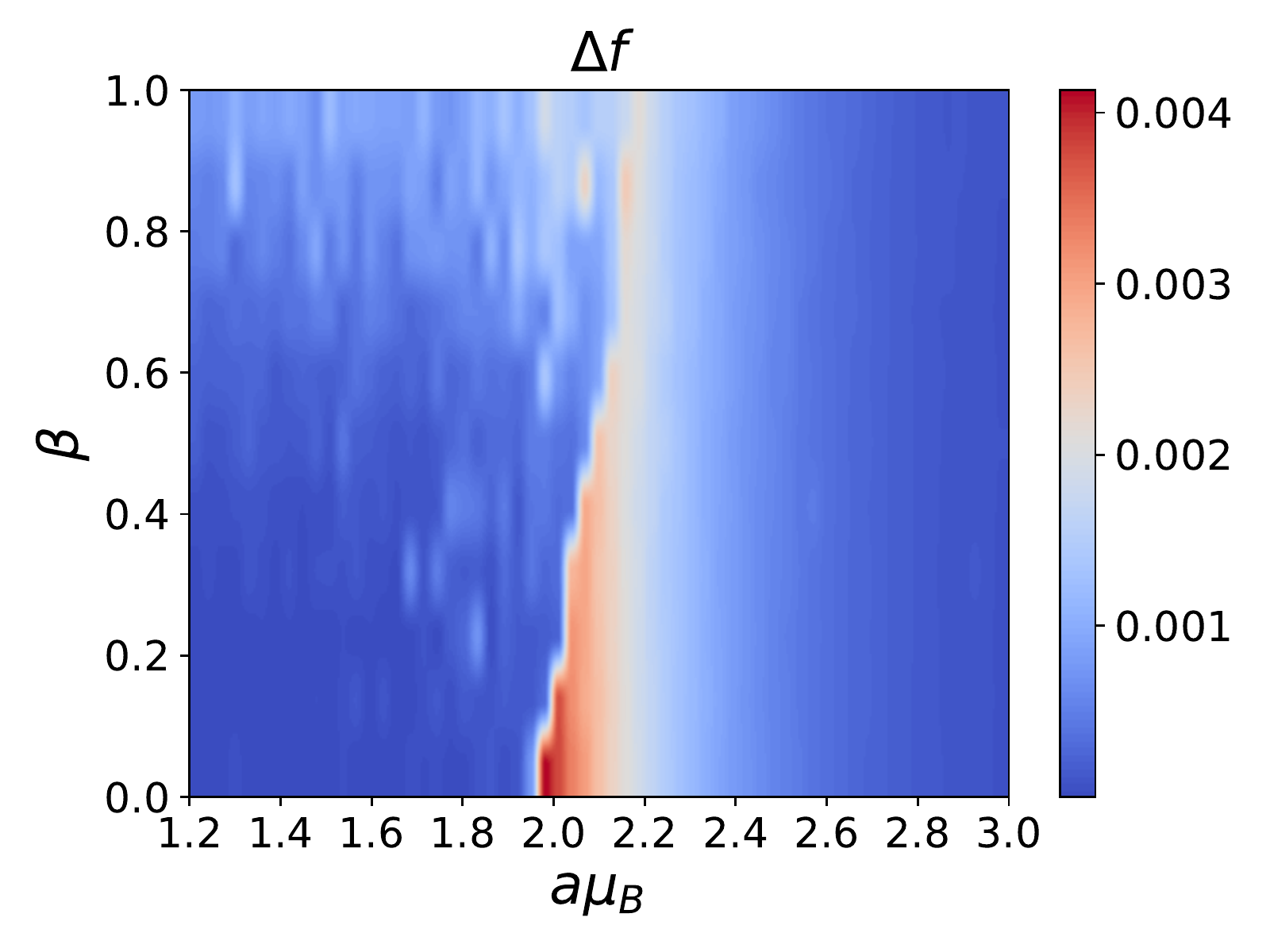}
\caption{$\Delta f$ at $am_q=0.2$ as a function of chemical potential and $\beta$ the on a $6^3 \times 4$ lattice}
\label{SignOBeta}
\end{figure}

Whereas baryons are point-like in the strong coupling limit, they become resolved as fermions split around plaquettes with non-zero plaquette occupation number. The fermion world-lines no longer have simple geometries and introduce an additional sign problem. This is shown in \fref{SignOBeta}:
up to $\beta=1$ the sign problem is still manageable: the nuclear transition weakens, however the sign problem gets more severe for all values of the chemical potential.  

\subsection{Determination of the baryon mass}
\label{BaryonMass}

We have argued above that the onset of the nuclear transition is roughly of the order of the nucleon mass. Since we limit ourselves in this paper to one staggered flavor, and there is no distinction between protons and neutrons (but see \cite{Unger2022} for the $\Nf=2$ generalization), we refer to the nucleon mass as baryon mass. We will see below that it is strongly quark mass dependent.

While it is not straight forward to measure the pion mass (it requires temporal extends larger than $\Nt=8$ and has been studied only in the continuous time limit by one of us \cite{Klegrewe2020}

In contrast to the 3-dim.~effective theory with Wilson fermions \cite{Langelage2014}, we also can not rely on a joint hopping parameter and character expansion to obtain analytic predictions for the baryon mass. Hence we determine it numerically. 

In the dual formulation, a closed baryon loop in the temporal direction is called a static baryon which has links $\bar{B}(x)B(y)$. This is a baryon hopping where a baryon is annihilated at site $x$ and created at site $y$. The probability for the hopping depends on the baryon mass $am_B$ and is proportional to $e^{-am_B}$ at low temperatures ($aT=\frac{1}{N_t}$). Therefore, the probability for having a static baryon loop with $\Nt$ links is proportional to $e^{-am_BN_t} = e^{\frac{-m_B}{T}}$.\\

We can express this probability as $e^{\frac{-\Delta F}{T}}$ with $\Delta F$ being the difference in free energy between configuration with a static baryon and configuration without. Equating this with the previous expression for the probability, we have $m_B=\Delta F$. This has been used to calculate the baryon mass previously \cite{Fromm2010}. The free energy ($F=E-TS$) is approximately equal to the energy at low temperatures which allows us to calculate the baryon mass with $\Delta E$ instead, which is better suited to cover the whole quark mass range. \\

Observables like the energy density can be numerically determined as an expectation values of dual variables, for $\Nc=3$:
\begin{align}
a^{4} \epsilon= \frac{\xi}{\gamma} \frac{\partial \gamma}{\partial \xi}\left\langle 2 N_{D t}+3 N_{B t}\right\rangle - \left\langle N_M \right\rangle
\end{align}
with $N_{Dt}$ the number of temporal dimers, $N_{Bt}$ the number of temporal baryon segments and $N_M$ the number of monomers. It also has a nontrivial dependence dependence on the bare anisotropy $\gamma$ and the physical anisotropy $\xi=a/a_t$ due to the derivative $\partial\gamma/\partial \xi|_{\gamma=1}$ even for isotropic lattices $\gamma=1$. We have determined the relation between $\xi$ and $\gamma$ for various quark masses \cite{Bollweg2018} to determine the derivative.\\

\begin{figure}[h!]
  \centering
\includegraphics[width=\columnwidth]{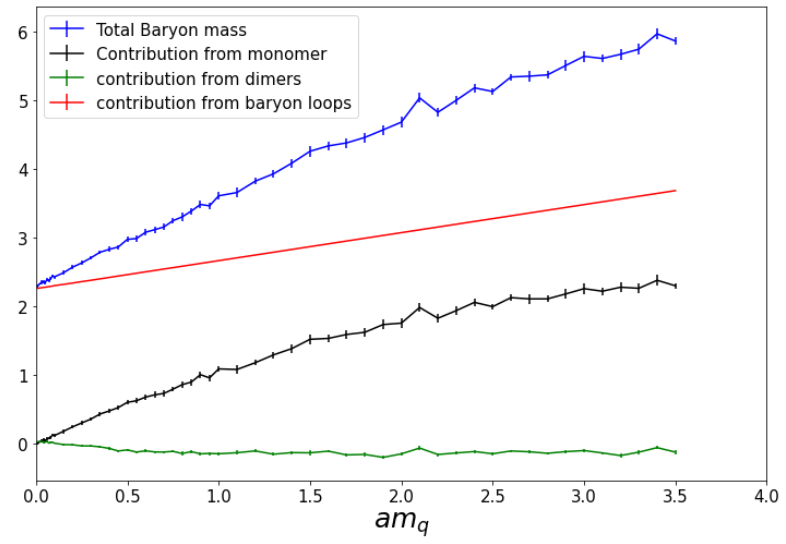}
\caption{Baryon mass from $\Delta E$ as a function of the quark mass $am_q$, and contributions from different dual variables: monomers, dimers and baryon segments.}
\label{Baryon_mass}
\end{figure}

Using the expression for energy density, we have calculated the baryon mass from the energy difference of an ensemble with and without a static baryon at the origin. Simulations of both ensembles were obtained via the  worm algorithm on a $8^3 \times 8$ lattice, i.e.~for $aT=0.125$, which is low enough to mimic zero temperature. The baryon mass is plotted as a function of the bare quark mass $am_q$ in Fig.\ref{Baryon_mass}. It is found that the baryon mass increases drastically with the quark mass.
This figure also show the contributions from different dual degrees of freedom on the baryon mass: whereas the main contribution in the chiral limit stems from the static baryon (with a minor contribution from temporal dimers), the main contribution for large quark masses is due to the monomers, whereas the contribution from dimers turn negative. 

As we did not find a strong $\beta$-dependence concerning the nuclear transition, it is well justified that the baryon mass will also not depend much on $\beta$, and we will benchmark all results with this strong coupling baryon mass.

\subsection{Determination of the nuclear transition}

\subsubsection{Strong coupling}

For low quark masses at intermediate temperatures, the nuclear transition is established at the strong coupling limit \cite{Kim2016}. In this paper, we have extended the results to larger quark masses and lower temperatures. In this region the nuclear critical end point is at much larger values of $a\mu_B$. 
All simulations are carried out on isotropic lattices with $aT=0.125$. This temperature is low enough to have an approximate silver blaze property, i.e. up to the nuclear transition, all observables are independent of $a\mu_B$.

\begin{figure}[h!]
\centering
\includegraphics[width=\columnwidth]{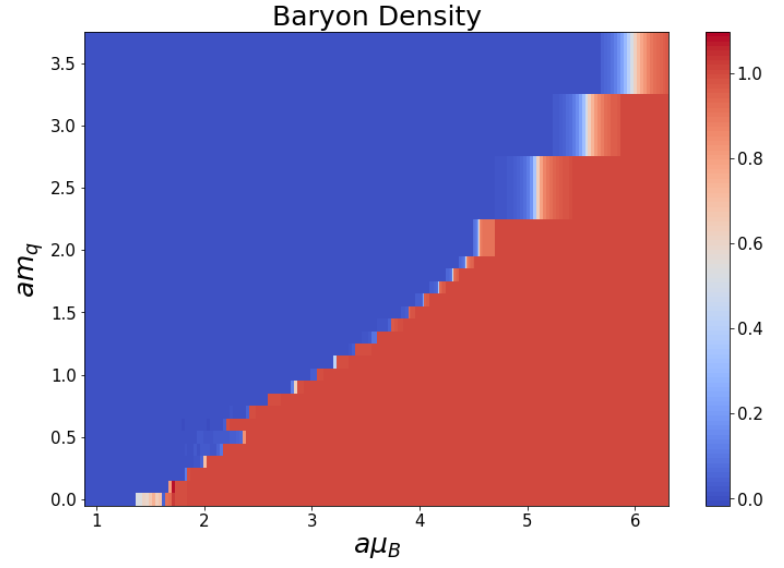}
\caption{Baryon density for volume $4^3\times8$ in the full $\mu_B-m_q$ plane, illustrating the strong quark mass dependence of the onset to nuclear matter.}
\label{BDensContour}
\end{figure}

\begin{figure*}[h!]
\centerline{
\includegraphics[width=0.9\columnwidth]{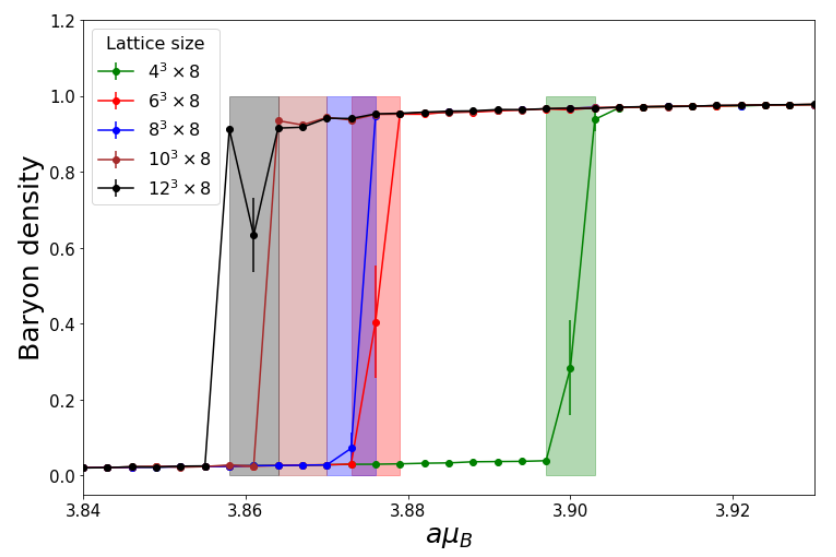}\quad 
\includegraphics[width=0.9\columnwidth]{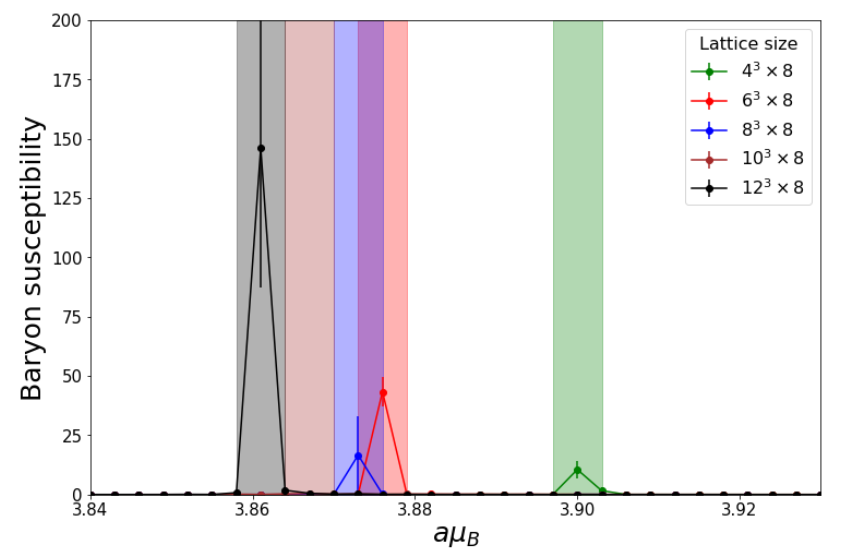}}
\centerline{\includegraphics[width=0.9\columnwidth]{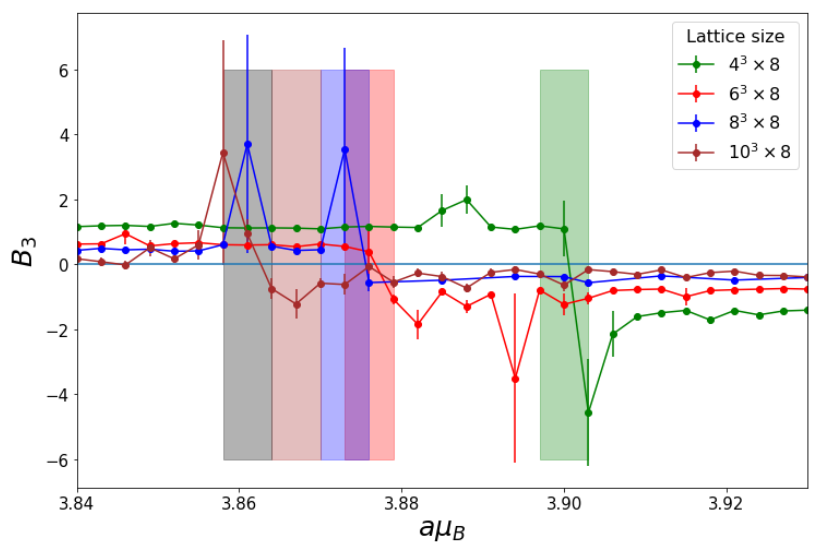}\quad
\includegraphics[width=0.9\columnwidth]{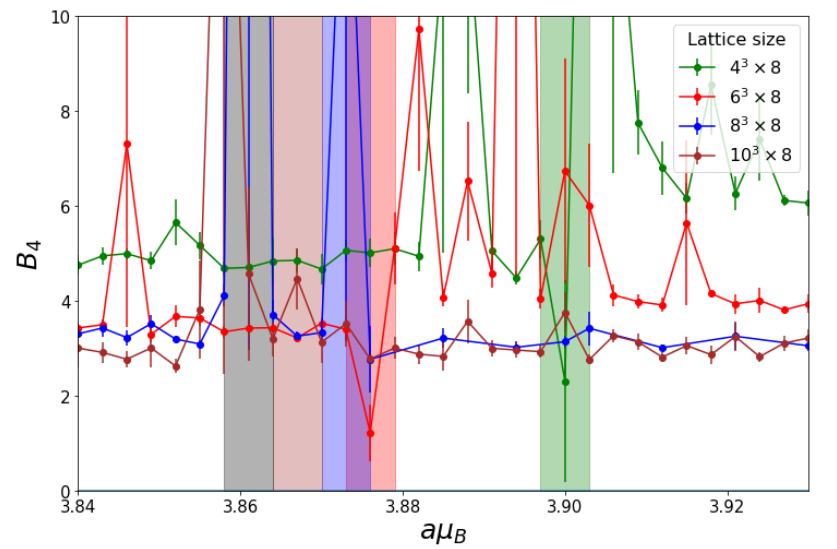}}
\caption{Baryonic observables on various volumes in the first order region $am_q=1.5$. Vertical bands indicate the mean and error of the nuclear transition.}
\label{BDensFirst}
\end{figure*}

\begin{figure*}[h!]
\centerline{
\includegraphics[width=0.9\columnwidth]{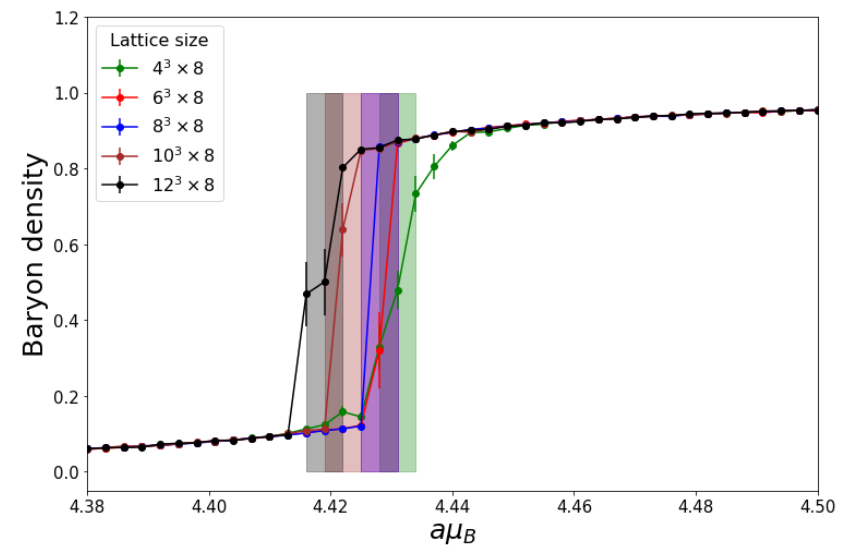}\quad 
\includegraphics[width=0.9\columnwidth]{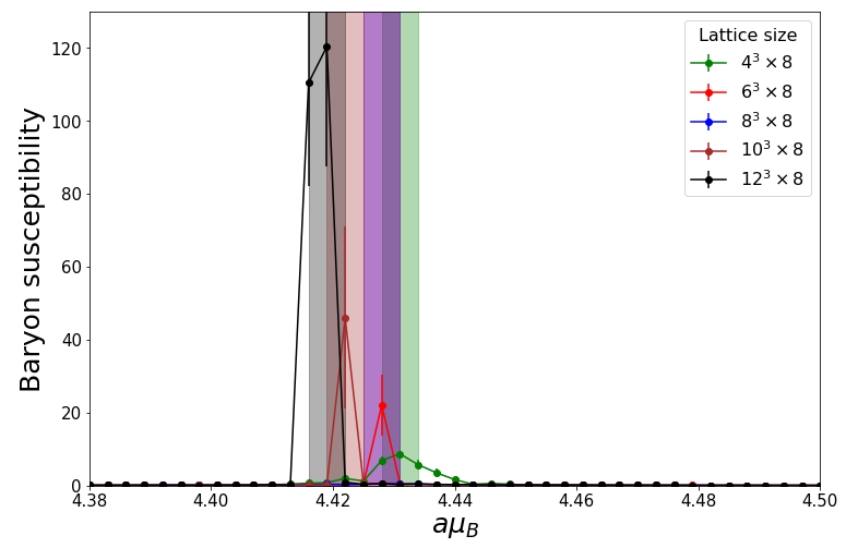}}
\centerline{\includegraphics[width=0.9\columnwidth]{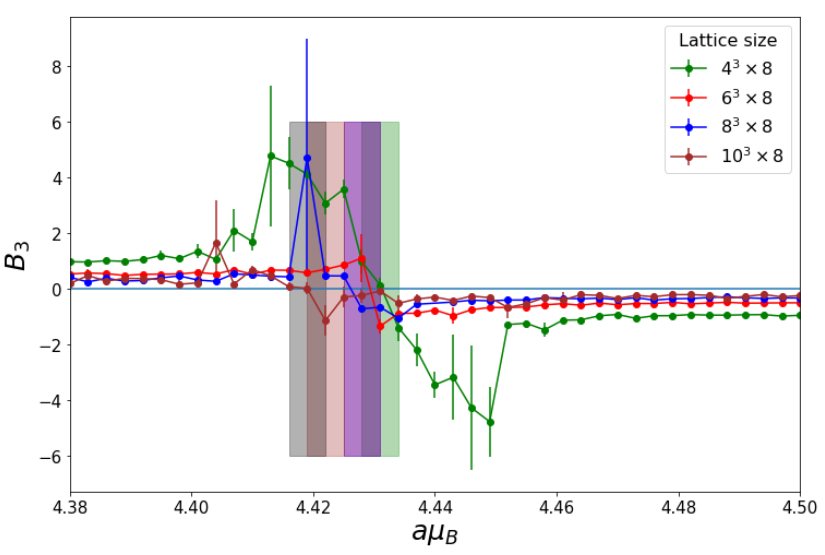}\quad
\includegraphics[width=0.9\columnwidth]{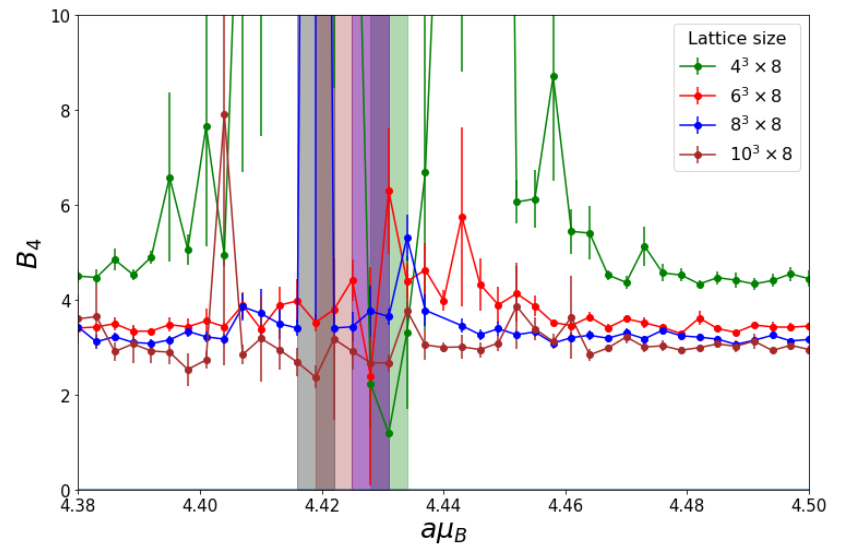}
}
\caption{Baryonic observables on various volumes in the crossover region $am_q=1.9$. Vertical bands indicate the mean and error of the nuclear transition.}
\label{BDensCross}
\end{figure*}

\begin{figure*}
\centerline{
\includegraphics[width=0.9\columnwidth]{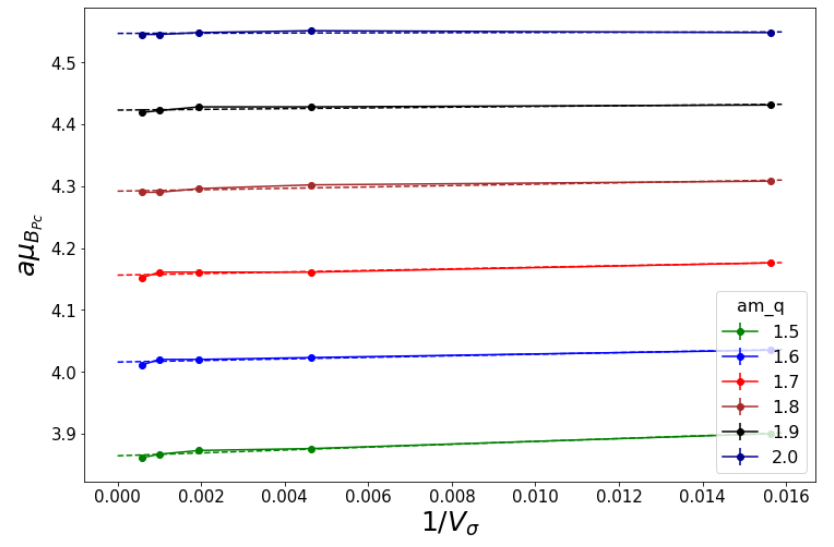}\quad
\includegraphics[width=0.9\columnwidth]{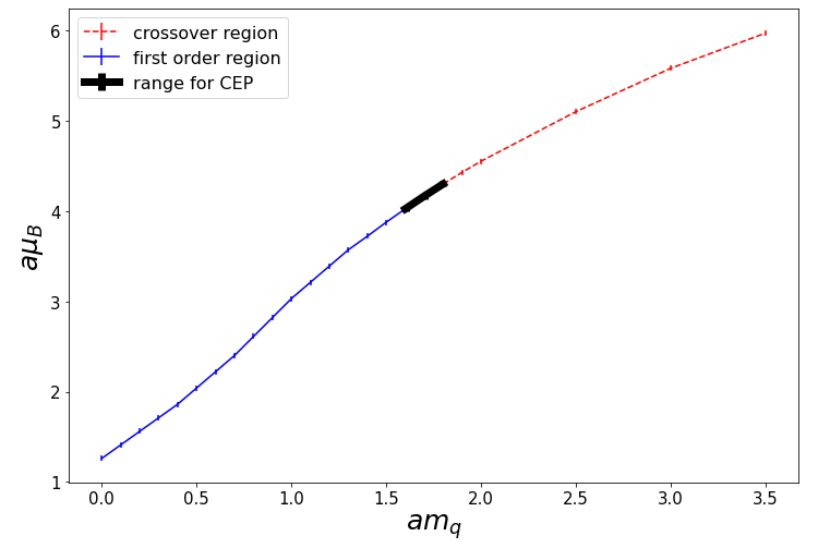}}
\caption{\emph{Left:} Extrapolation of the pseudo-critical values of $\mu_B$ for the various volumes into the thermodynamic limit. \emph{Right:} Critical baryon chemical potential for different quark masses. The first order transition region is shown in blue, the crossover region is shown in red and the range for critical end point is marked in black.}
\label{CEP}
\end{figure*}

With the dual formulation, we can calculate the baryon density and the baryon susceptibility as a function of the baryon chemical potential for different quark masses. For a first order transition, the baryon density shows a discontinuity as shown in \fref{BDensFirst} for $am_1=1.5$. Note that the resolution is high enough to obtain $a\mu_B^{1st}$ to high accuracy, the error in $a\mu_B^{1st}$ is mainly due to this resolution $\Delta a\mu_B=0.003$ (see \tref{SCTable}). However, since the first order transition is strong, not all peaks of the susceptibility could be resolved. The skewness $B_3$ becomes zero at the transition, and the Binder cumulant gets close to the first order value $B_4=3$.
At sufficiently large larger quark masses, this transition becomes continuous (a rapid crossover) as seen in \fref{BDensCross} for $am_q=1.9$. For a crossover transition there is no well defined value for $\mu_{Bc}$, so we take the critical chemical potential to be the value at which the baryon density is $n_B=0.5$.  We find that the first order transition has a smaller gap beyond quark mass $am_q=1.5$ and it has vanished at $am_q=1.8$. 
In \fref{BDensContour} we show the baryon density in the full $\mu_B-m_q$ plane, where it can seen how the transition broadens in the crossover region.
We have extrapolated the pseudo-critical chemical potentials to the thermodynamic limit to obtain the phase boundary and the order of the transition as a function of the quark mass, as shown in \fref{CEP}.
The best guess for the nuclear critical end point at $aT=0.125$ is $am_q^c=1.7(1)$, which corresponds to $am_\pi^c\simeq 3.10$ or $m_\pi^c/m_B\simeq 0.64$.


\subsubsection{Finite $\beta$}

The extension to finite $\beta$ requires more statistics, and each simulation takes longer due to the inclusion of the plaquette occupation numbers. Hence we could only address the temperature $aT=0.25$ ($\Nt=4$), which is not low enough to find the silver blaze property as in the strong coupling limit. As a consequence, the nuclear transition is weaker and the values of $am_B^{1st}$ and $am_B^{c}$ will not coincide with the lower temperature counterpart. But here we will now get insight into the $\beta$-dependence of the nuclear transition. 

\begin{figure*}[h!]
\centerline{
\includegraphics[width=0.9\columnwidth]{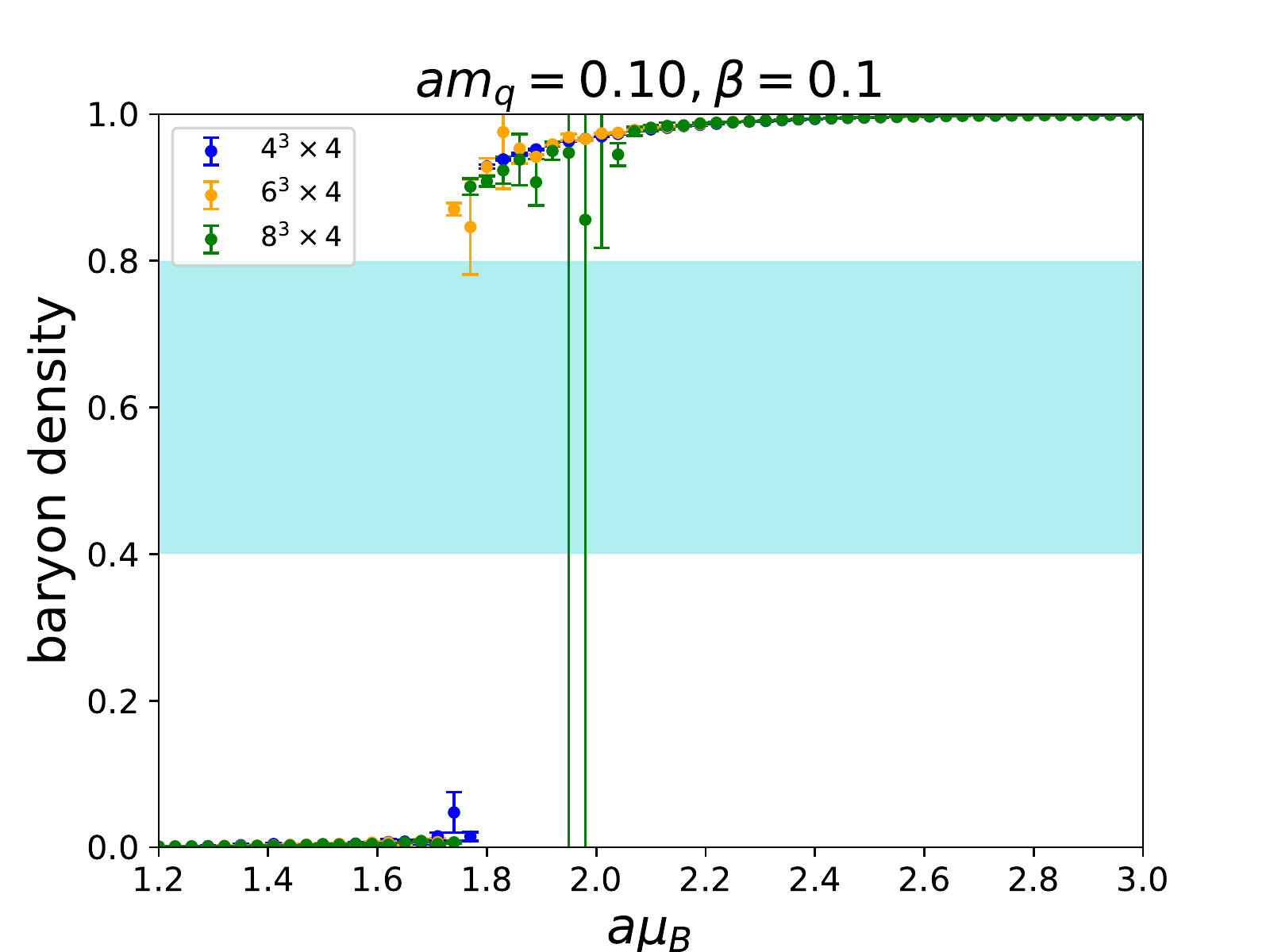}\quad
\includegraphics[width=0.9\columnwidth]{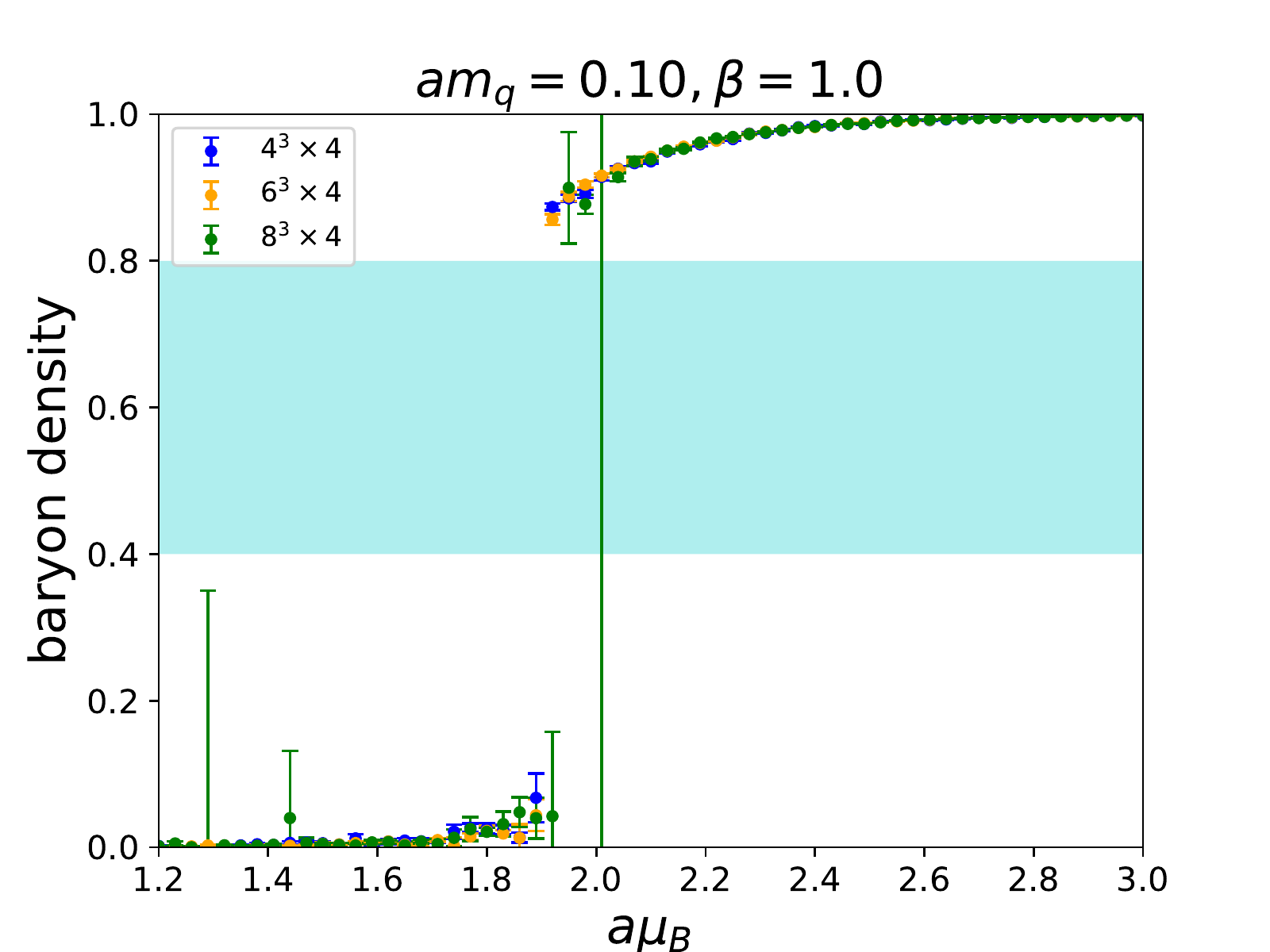}
}
\centerline{
\includegraphics[width=0.9\columnwidth]{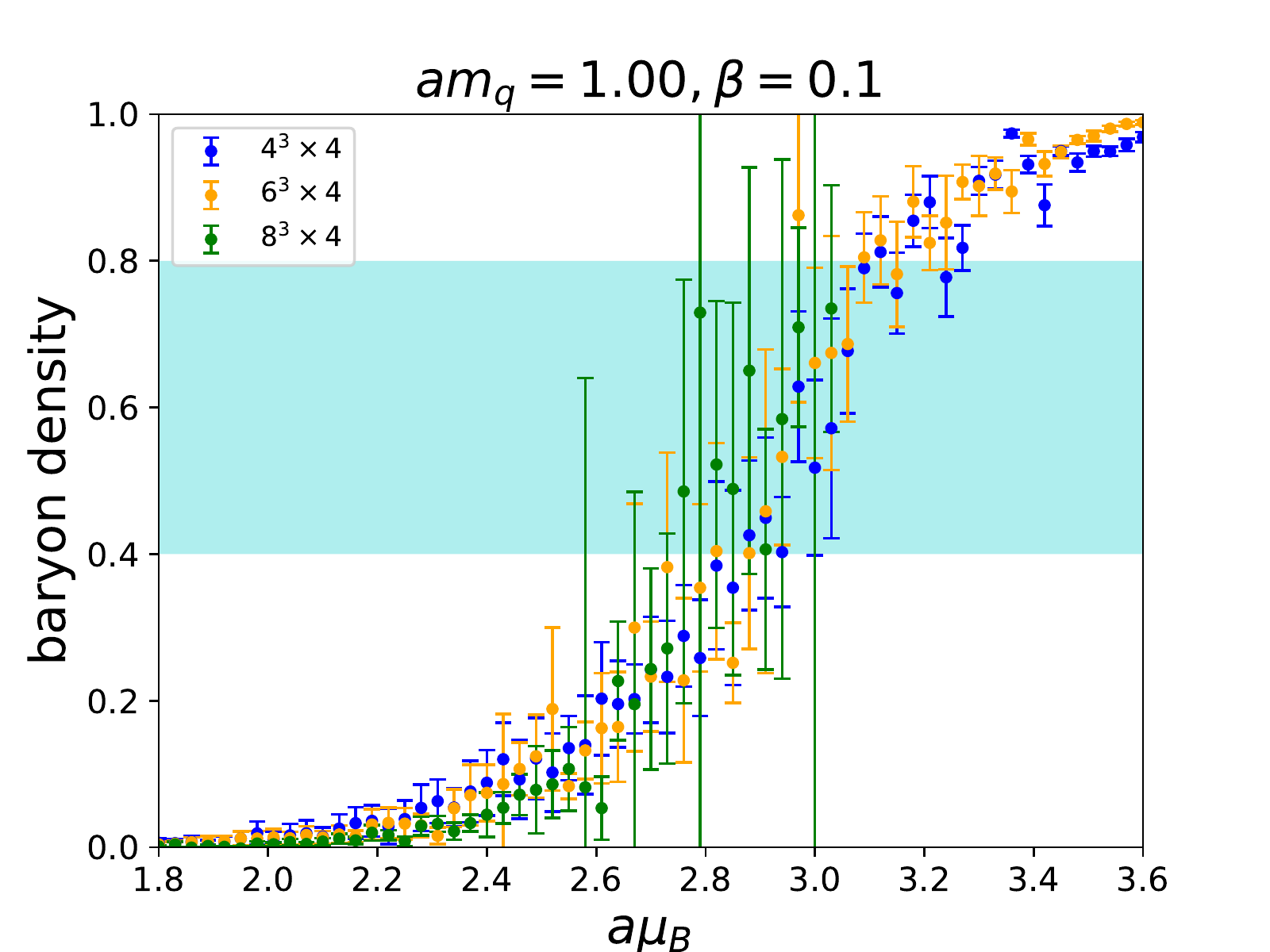}\quad
\includegraphics[width=0.9\columnwidth]{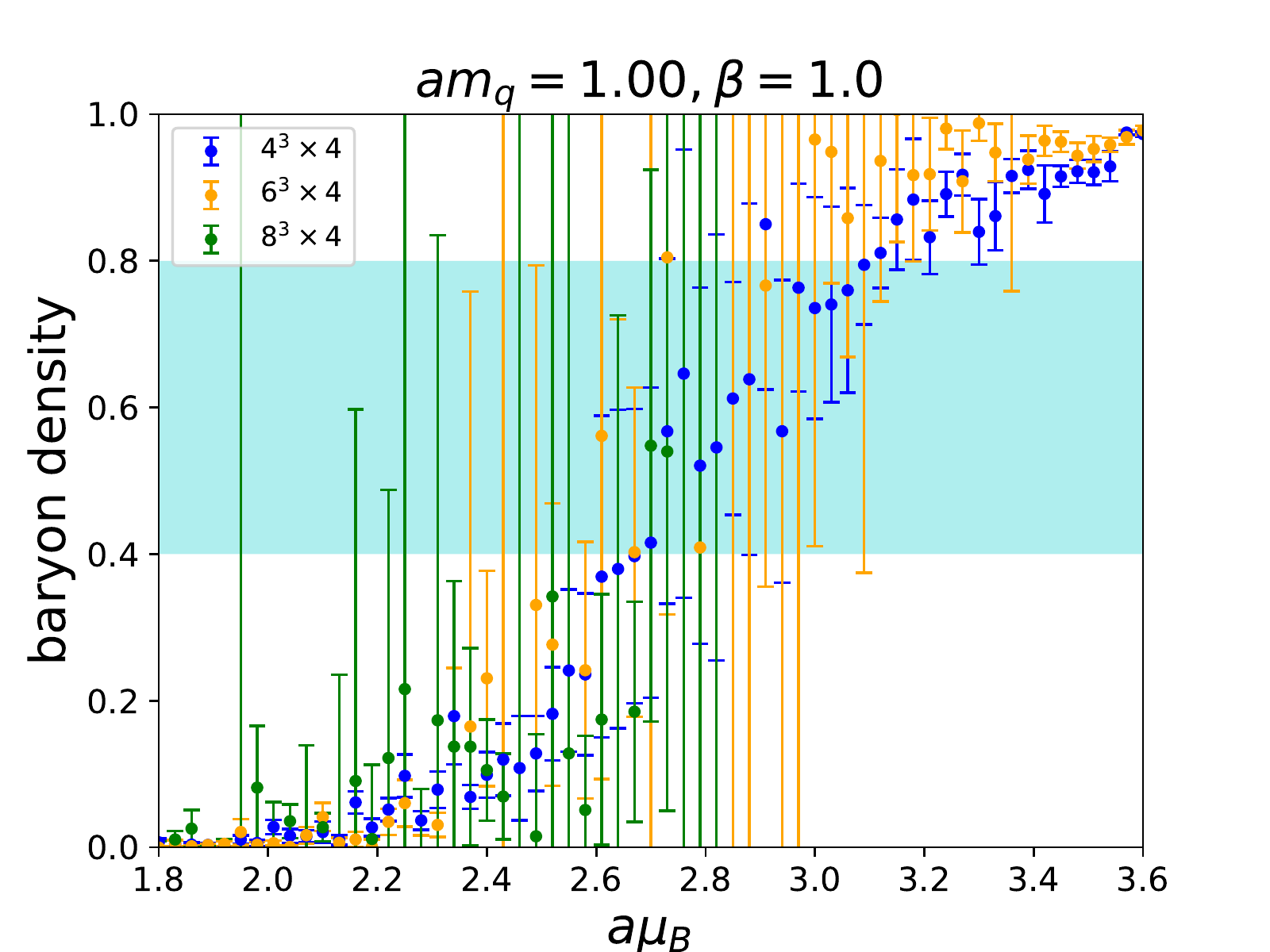}
}
\caption{The baryon density at small quark mass 
$am_q=0.1$ \emph{(top)} and large quark mass $am_q=1.0$ \emph{(bottom)}, and small inverse gauge coupling $\beta=0.1$ \emph{(left)}
and a large value $\beta=1.0$ \emph{(right)}.
The $\beta=0.1$ results still show a strong first order transition for both quark masses, which turns into a crossover for $\beta=1.0$. 
The blue bands indicate how the error on the onset $a\mu_B^{pc}$ was obtained, which is particularly relevant in the crossover region.
}
\label{BDensOB}
\end{figure*}

In \fref{BDensOB} we show the baryon density as a function of the baryon chemical potential for two sets of quark masses $am_q=0.1$, $am_q=1.0$ and inverse gauge couplings $\beta=0.1$, $\beta=1.0$.
For small quark masses, there is a strong first order phase transition. It is slightly weaker for $\beta=1.0$ compared to $\beta=0.1$, but still strong with roughly the same value of 
$\mu_B^{1st}=1.85(5)$. 
At large quark masses however, the transition is crossover. 
The blue bands in the figures are used to determine the pseudo-critical baryon chemical potentials  $a\mu_B^{pc}$ for each parameter in each finite volume. We determine the uncertainty of the critical baryon chemical potential from the range of the data points below/above the blue band. Since the shape of the transition is asymmetric at finite temperature, we chose the band also asymmetric, ranging from $n_B=0.4$ to $n_B=0.8$.





\begin{figure*}[h!]
\centering
\includegraphics[width=\columnwidth]{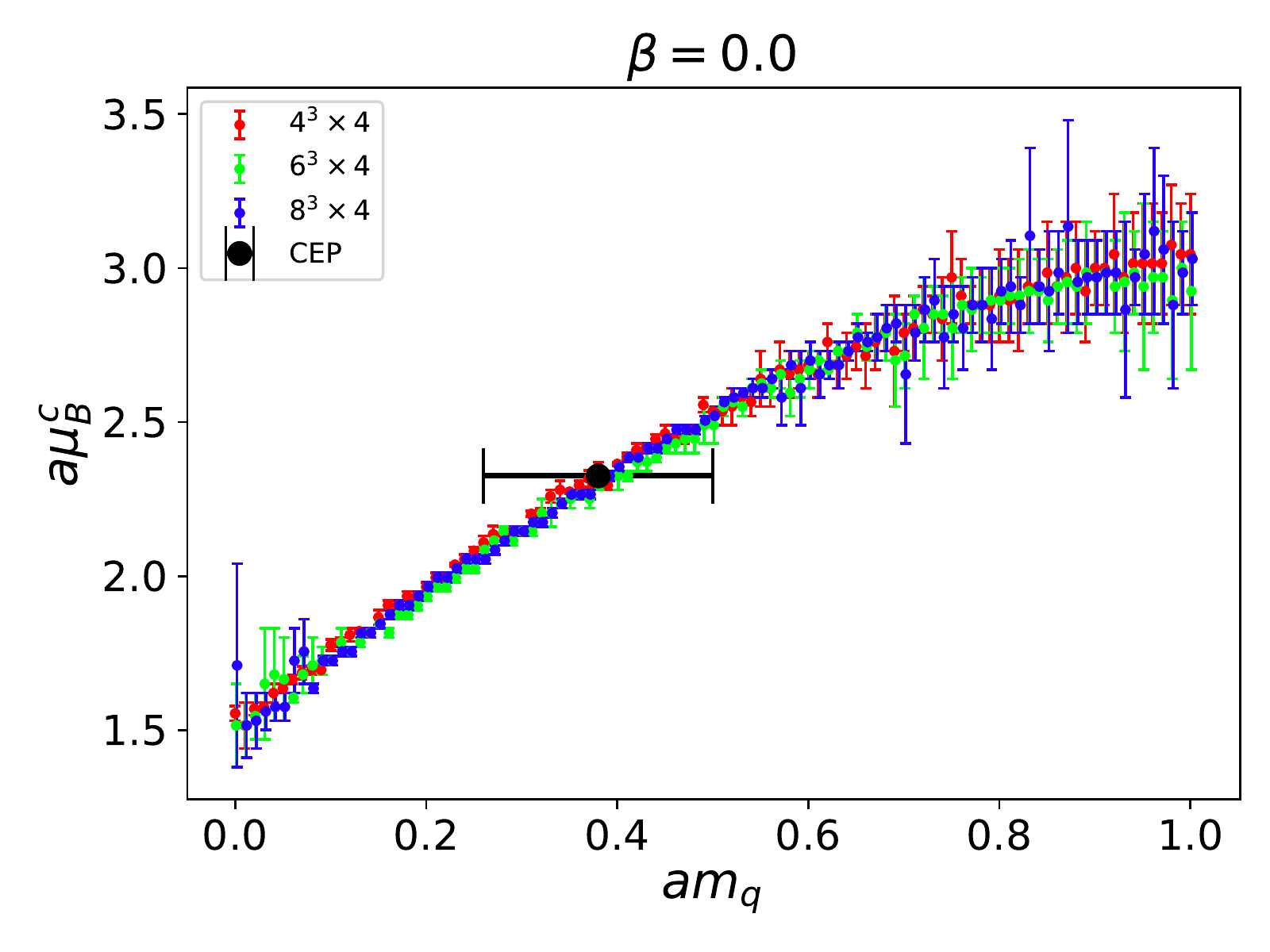}
\label{fig:mu_cri_beta_0.0}
\centering
\includegraphics[width=\columnwidth]{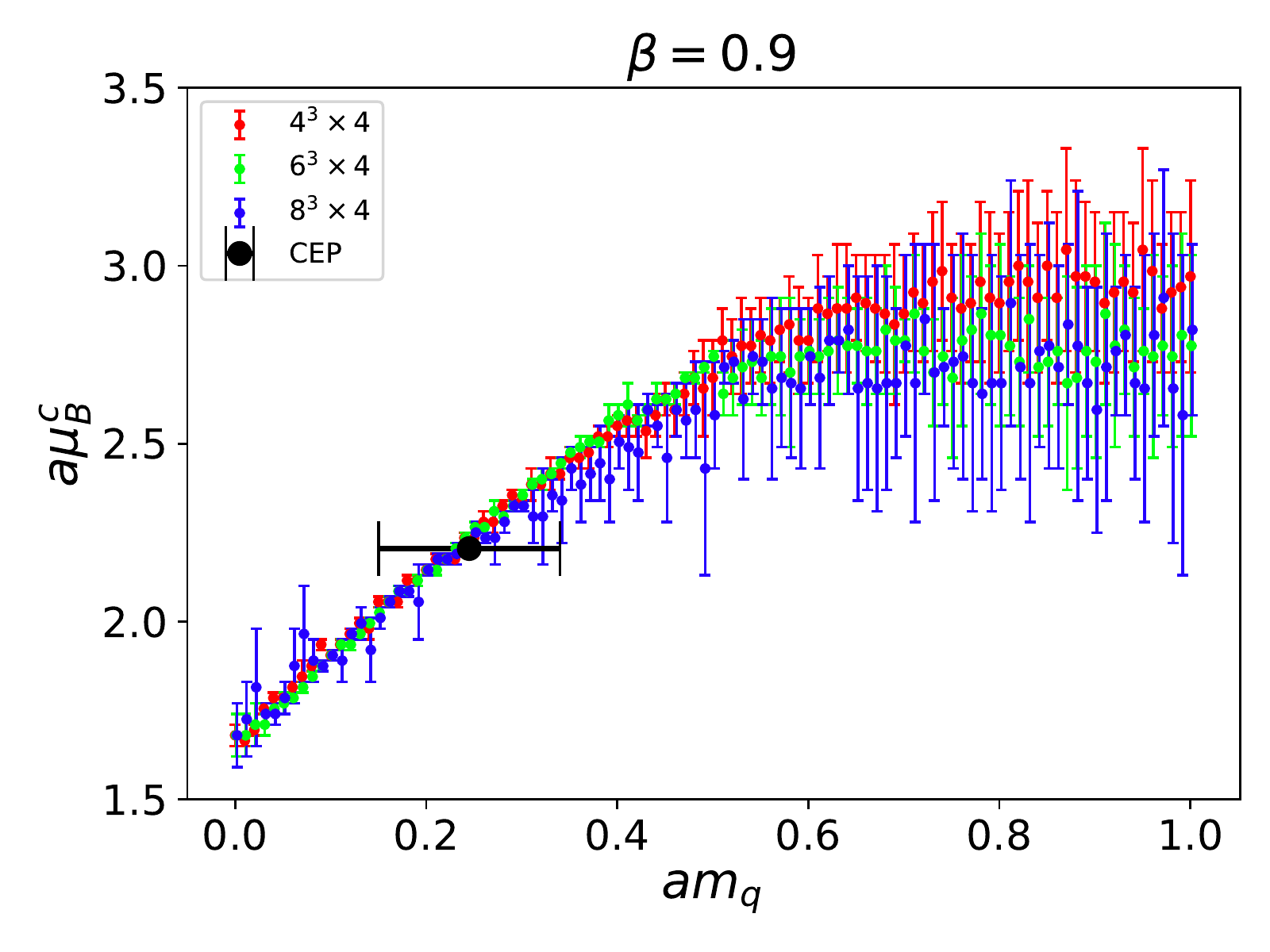}
\caption{The critical baryon chemical potential as a function of quark mass at $\beta=0$ (\emph{left}) and at $\beta=0.9$ (\emph{right}) for 3 different volumes and a range for the location of the critical end point, which was obtained from the histogram analysis as shown in \fref{HistogramBDens}.}
\label{NuclearBetaMass}
\end{figure*}

The quark mass dependence is shown in \fref{NuclearBetaMass} for two values, $\beta=0.0$ and $\beta=0.9$. While the location of the critical point moves to lower quark masses at larger values of $\beta$, there is only a very weak dependence of the actual phase boundary on $\beta$. Essentially, the quark mass dependence of the phase boundary is the same.

\begin{figure*}[h!]
\centerline{
\includegraphics[width=0.33\textwidth]{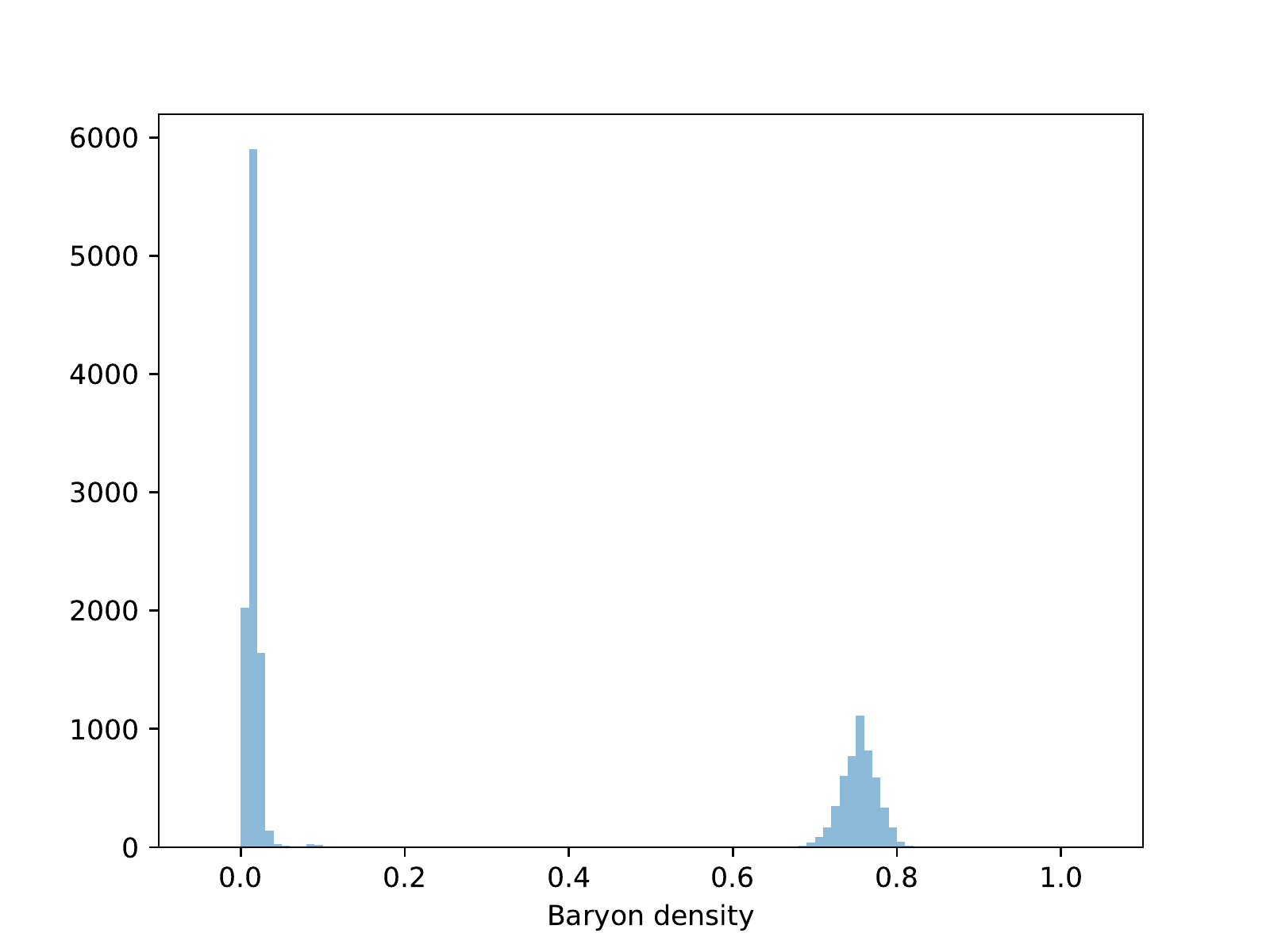}\;
\includegraphics[width=0.33\textwidth]{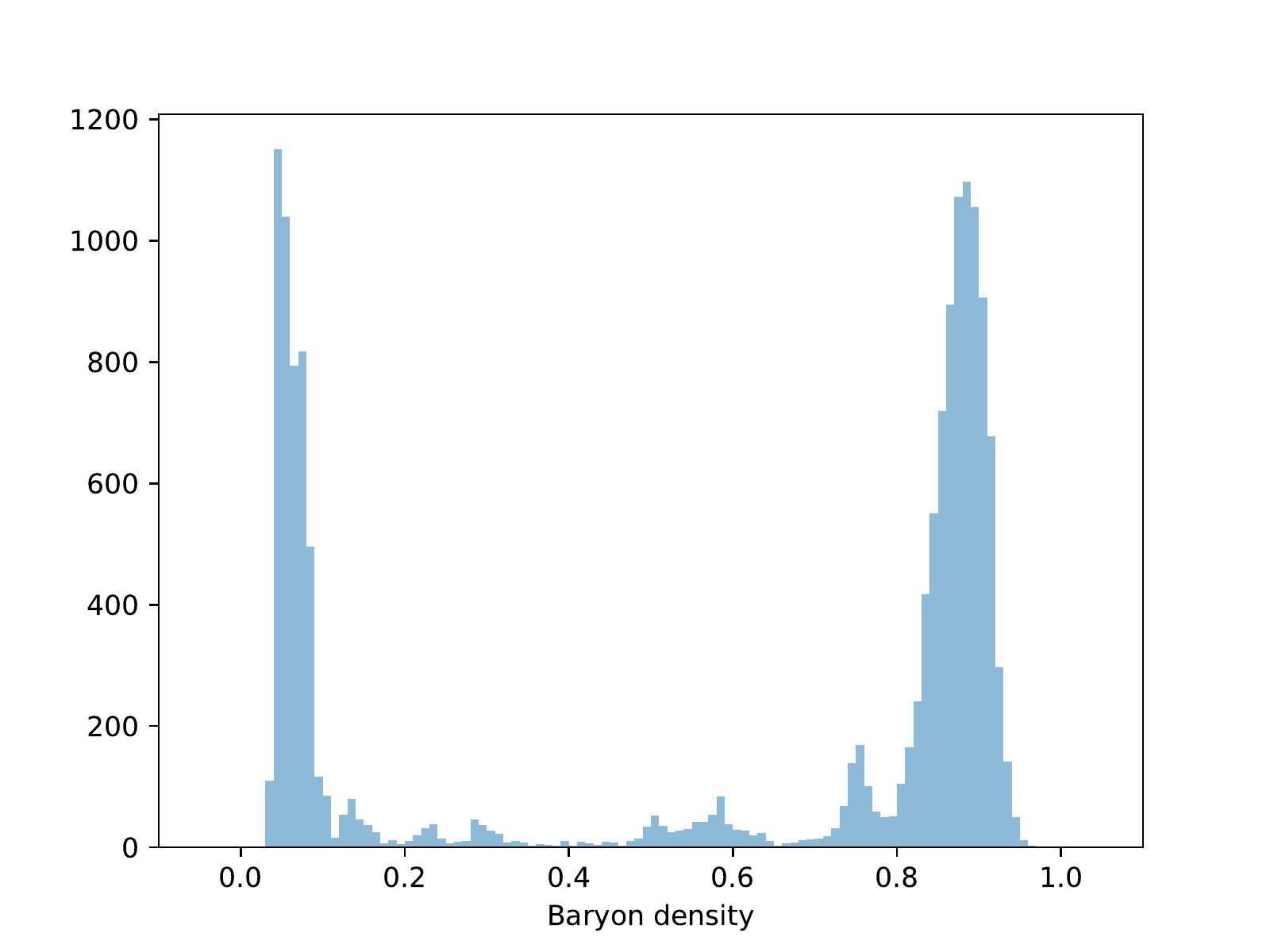}\;
\includegraphics[width=0.33\textwidth]{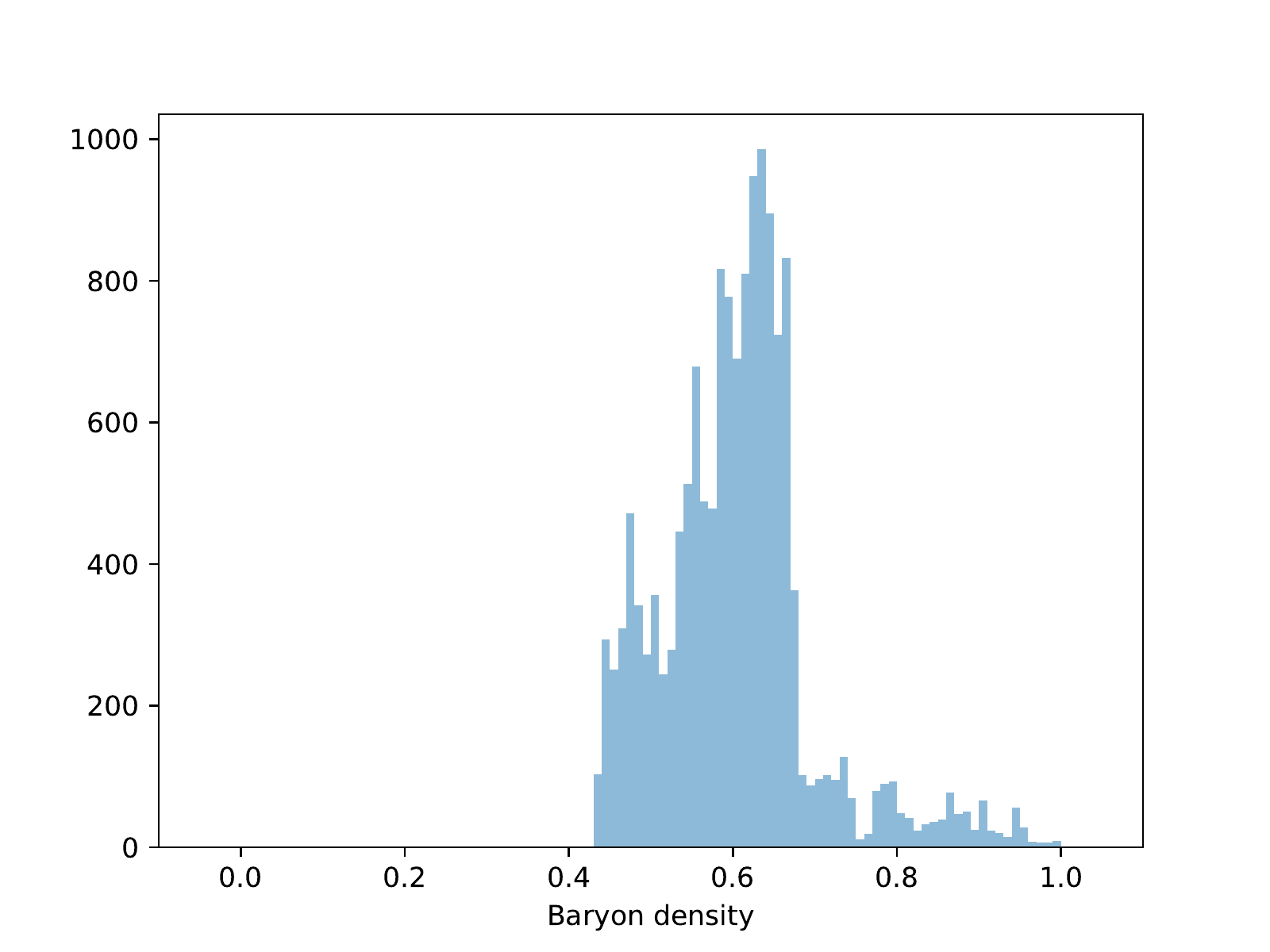}
}
\caption{Histogram of baryon density for various parameters: \emph{Left:} at the first order phase transition with two peaks ($\beta=0.9$, $am_q=0.02$, $a\mu_q=0.559$).
\emph{Center:} at the second order phase transition with two peaks and small bump between them at ($\beta=0.9$, $am_q=0.24$, $a\mu_q=0.748$).
\emph{Right:} at the crossover transition with one peak in the middle at ($\beta=0.9$, $am_q=0.48$, $a\mu_q=0.926$).}
\label{HistogramBDens}
\end{figure*}

\begin{figure*}[h!]
\centerline{
\includegraphics[width=0.98\columnwidth]{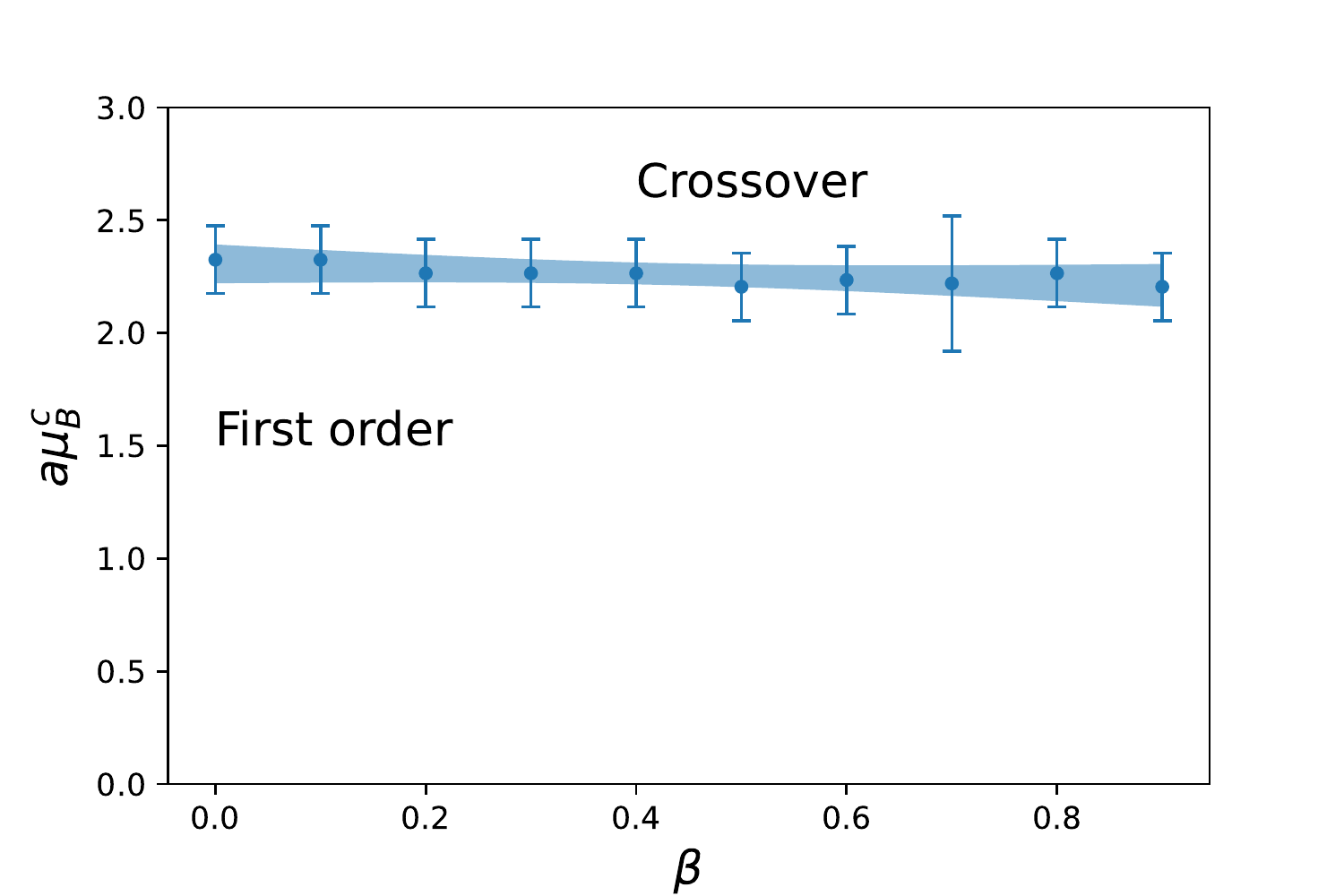}\quad
\includegraphics[width=0.88\columnwidth]{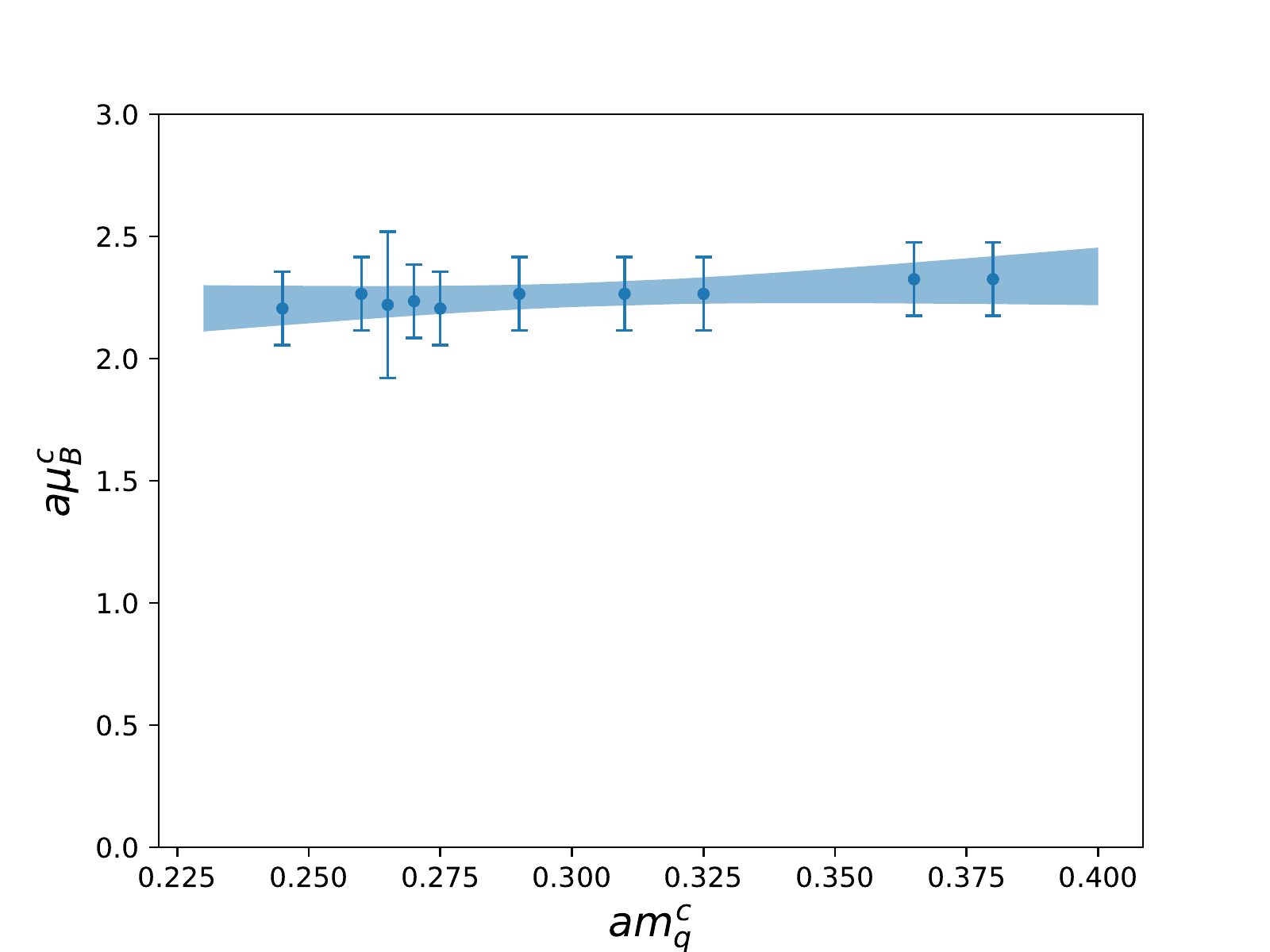}
}
\caption{The critical baryon chemical potential $a\mu_B^c$ as a function of $\beta$ (\emph{left})
and the critical quark masses $am_q^{c}$ (\emph{right}).}
\label{fig:endpoints}
\end{figure*}

More information on the intermediate range between the strong first order transition at small $\beta$ and the crossover at larger $\beta$ can be obtained by a histogram analysis: If the transition is first order, then at $a\mu_B^{1st}$ there should be two peaks in the histogram of baryon density. 
As the quark mass increases, the transition eventually becomes second order, and a small bump appears at intermediate baryon density. At large quark mass, a broad peak forms in the middle which indicates crossover transition. This is shown in \fref{HistogramBDens}. The values of $a\mu_B$ can be fine-tuned to yield a histogram which has approximately same peak height at low and high density. From this, $a\mu_B^{1st}$ can be reliably extracted, and an estimate of the critical point  $a\mu_B^{c}$ obtained. This is shown in \fref{fig:endpoints}.
Note that this analysis was not possible at strong coupling as the very low temperature did not yield histograms with enough structure. 

\clearpage

\subsection{Nuclear interactions}

We have found that $a\mu_B^{1st}$ is very different from the baryon mass. This must be due to strong attractive interactions of nucleons. In contrast to continuum physics, in the strong coupling limit there is no pion exchange due to the Grassmann constraint. Instead, nucleons are point like and hard core repulsive. However, the pion bath, which is modified by the presence of static baryons, results in an attractive interaction. In \cite{Fromm2009}, this has been analyzed in the chiral limit using the snake algorithm, and it has been found that the attractive force is of entropic origin.

\begin{figure}[h!]
  \centering
\includegraphics[width=\columnwidth]{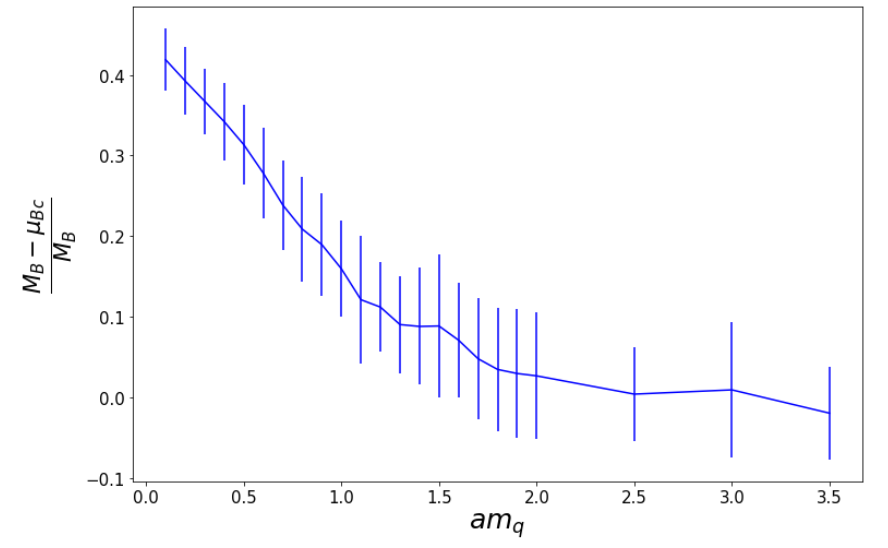}
\caption{Nuclear interaction scaled with baryon mass. As the quark mass increases, it tends to zero.}
\label{interaction}
\end{figure}

Here, we do not quantify the nuclear interaction via the nuclear potential, but via the difference between critical baryon chemical potential and baryon mass, in units baryon mass, as shown in \fref{interaction}, given the $am_B$ as measured in \sref{BaryonMass}. This compares better to the 3-dim.~effective theory.
The nuclear interaction is maximal and more than 40\% in the chiral limit, which is related to pions being massless: the modification of the pion bath is maximal. 
We clearly find that the nuclear interaction decreases drastically and almost linearly until it almost approaches zero at about $am_q=2.0$, corresponding to a pion mass $am_\pi=3.36$, see \sref{MF}. The large error bars for larger quark masses, that are due to the subtraction of almost same magnitudes, makes it difficult to extract a non-zero nuclear interaction at the largest quark masses. 

\section{Conclusion}

\begin{figure}[h!]
  \centering
\includegraphics[width=\columnwidth]{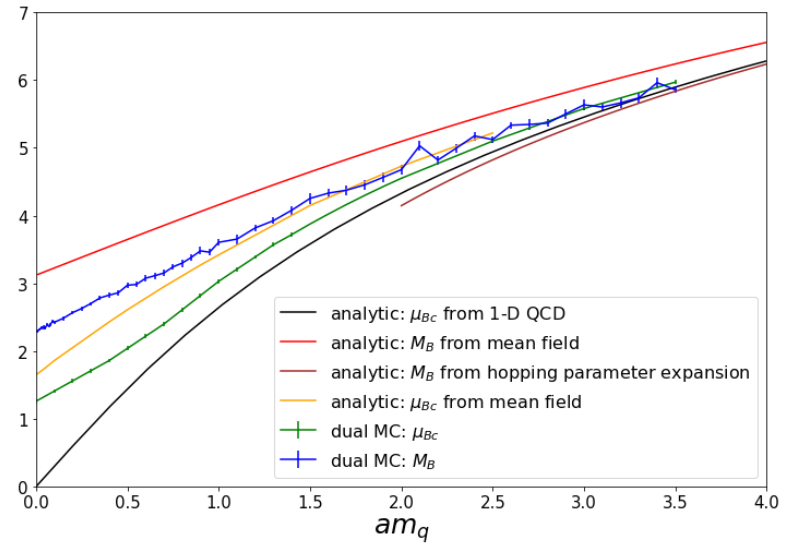}
\caption{Critical baryon chemical potential and baryon mass from different approaches.}
\label{comparison}
\end{figure}

In this work, we have determined the baryon mass and the nuclear transition via Monte Carlo: the worm algorithm based on the dual formulation, at finite $\beta$ equipped with additional updates. All those numerical results and various analytic expressions are summarized in \fref{comparison}. We find that as the quark mass becomes large, spatial mesons hoppings (i.e.~spatial dimers) become rare, which makes this 3+1-dimensional system closer to 1-dim.~QCD \cite{Bilic1988}. 
Also, both the baryon mass and the baryon chemical potential obtained in our dual representation, i.e.~for staggered fermions, approaches the baryon mass \cite{Hoek1982} of the 3-dim.~effective theory which is based on Wilson fermions.

\begin{figure}[h!]
 \centering
\includegraphics[width=\columnwidth]{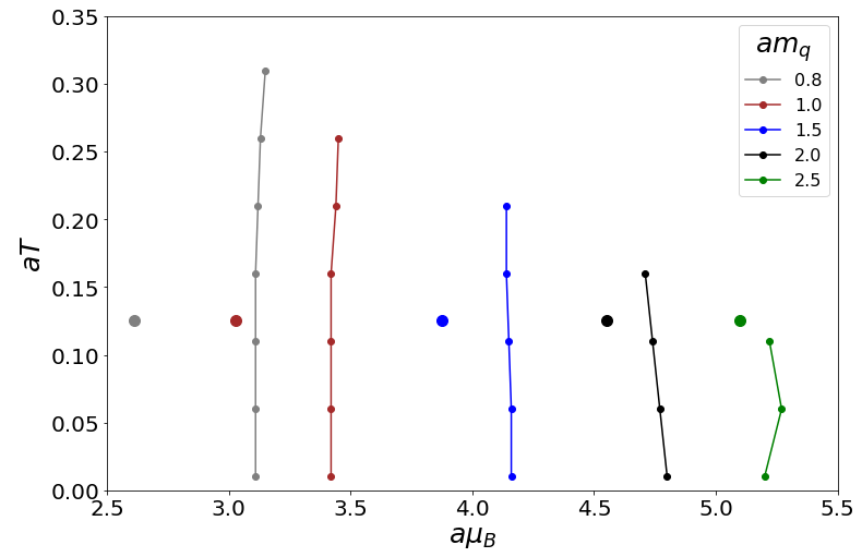}
\caption{Phase diagram obtained from the mean field approach and the critical chemical potential obtained from the dual formulation at strong coupling for $aT=0.125$. }
\label{phasediagMFcomapre}
\end{figure}

Another comparison that summarizes the validity of the mean field approach discussed in \sref{MF} is shown in \fref{phasediagMFcomapre}.
It is evident that mean field theory has strong deviations for small quark masses, but this discrepancy becomes smaller for larger quark masses.

\begin{figure}[h!]
\centering
\includegraphics[width=\columnwidth]{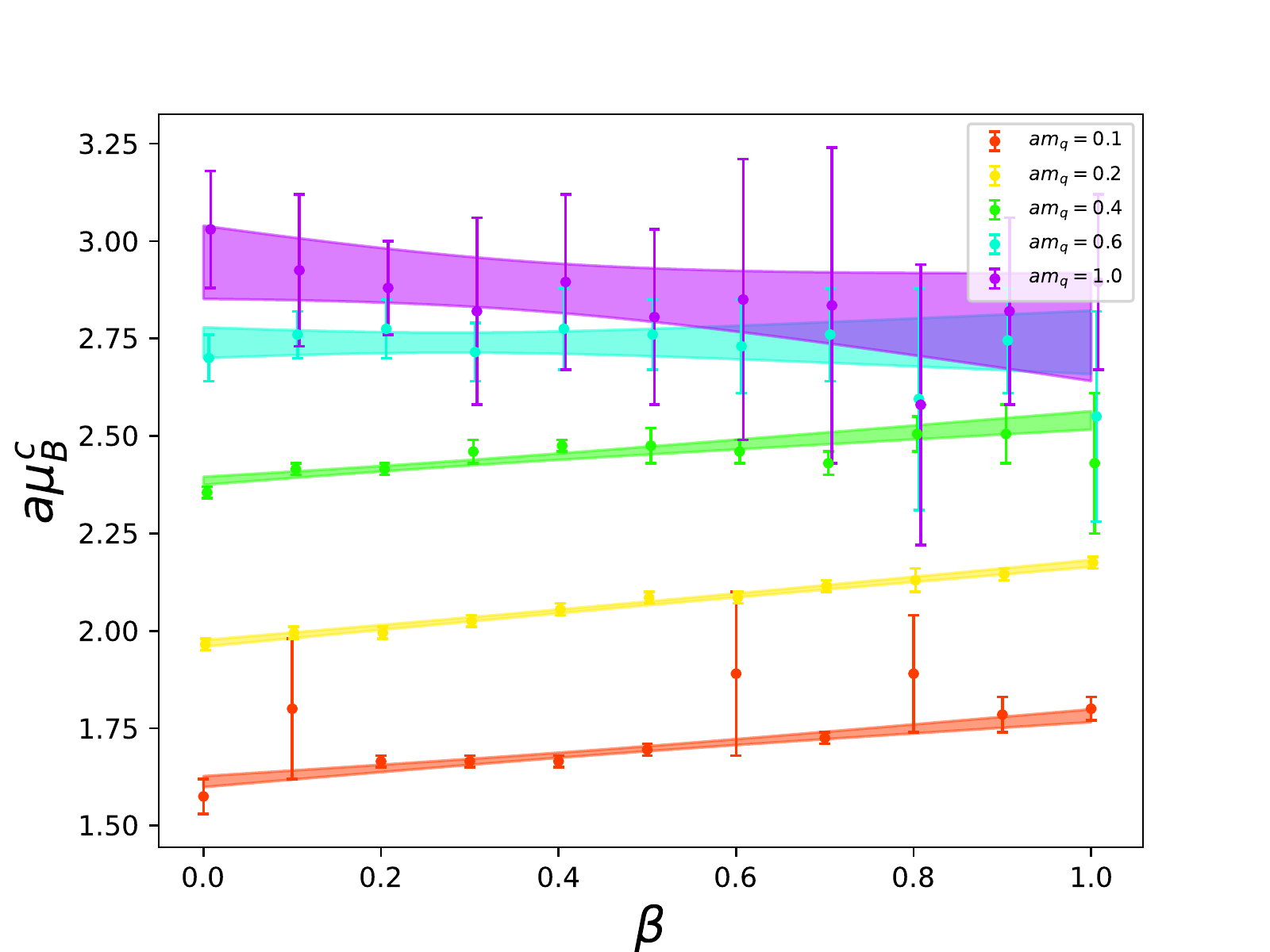}
\caption{The baryon chemical potential at fixed quark masses and linear fit as a function of $\beta$.}
\label{betafit}
\end{figure}

The extension of the study of the nuclear transition to finite inverse gauge coupling $\beta$ is summarized in \fref{betafit}, which shows the $\beta$-dependence of $a\mu_B^c$ for various quark masses. For all quark masses ranging from $am_q=0$ to $am_q=1.0$, there is only a very weak $\beta$-dependence, confirming the expectation from mean field theory \cite{Miura2009}.

This works was restricted to isotropic lattices\linebreak $\xi=a/a_t=1$, i.e. we performed simulations at fixed temperature. Non-isotropic lattices are necessary to vary the temperature at fixed values of $\beta$. This requires to  include two bare anisotropies, $\gamma$ for the fermionic action and $\gamma_G$ for the gauge action.  Finite $\beta$ has only been studied by us in the chiral limit \cite{Gagliardi2017}. Clearly, it is interesting to study the location of the nuclear critical point also including higher order gauge corrections and at finite quark mass. Simulations including $\mathcal{O}(\beta^2)$ are under preparation. 

\begin{acknowledgments}
We would like to thank Owe Philipsen for helpful discussions concerning comparison with the Polyakov effective theory in the heavy-dense regime.
The authors gratefully acknowledge the funding of this project by computing time provided by the Paderborn Center for Parallel Computing (PC²). We would like to thank the Center for Scientific Computing, University of
Frankfurt for making their High Performance Computing facilities available. This work is supported  by the Deutsche Forschungsgemeinschaft (DFG) through the CRC-TR 211 'Strong-interaction matter under extreme conditions'– project number 315477589 – TRR 211. J.K. was supported in part by the NSFC and the Deutsche Forschungsgemeinschaft (DFG, German Research Foundation) through the funds provided to the Sino-German Collaborative Research Center TRR110 "Symmetries and the Emergence of Structure in QCD" (NSFC Grant No. 12070131001, DFG Project-ID 196253076 - TRR 110)
\end{acknowledgments}

\clearpage

\section{Appendix}

\subsection{Statistics}

\subsubsection{Strong Coupling}

All runs at strong coupling have been obtained for $\Nt=8$, which corresponds to a rather low temperature $aT=0.125$ compared to the value of the chiral transition $aT\simeq 1.54$. 

\LTcapwidth=\columnwidth

\begin{longtable}{|c|c|c|c|c|}
\hline
$am_q$                                      & Volume                       & $a\mu_q$ & Steps & Worm updates \\ \hline
\multirow{3}{*}{0.0}                       & $4^3$ $\times$ 8  & {[}0.3-0.7{]}          & 0.01  & $100\times 10^7$               \\ \cline{2-5} 
                                           & $6^3$ $\times$ 8  & {[}0.3-0.7{]}          & 0.01  & $100\times 10^7$               \\ \cline{2-5} 
                                           & $8^3$ $\times$ 8  & {[}0.3-0.7{]}          & 0.01  & $100\times 10^7$               \\ \hline
\multirow{3}{*}{0.1}                       & $4^3$ $\times$ 8  & {[}0.5-0.7{]}          & 0.01  & $100\times 10^7$               \\ \cline{2-5} 
                                           & $6^3$ $\times$ 8  & {[}0.5-0.7{]}          & 0.01  & $100\times 10^7$               \\ \cline{2-5} 
                                           & $8^3$ $\times$ 8  & {[}0.5-0.7{]}          & 0.01  & $100\times 10^7$               \\ \hline
\multirow{3}{*}{0.2}                       & $4^3$ $\times$ 8  & {[}0.5-0.8{]}          & 0.01  & $100\times 10^7$               \\ \cline{2-5} 
                                           & $6^3$ $\times$ 8  & {[}0.5-0.8{]}          & 0.01  & $100\times 10^7$               \\ \cline{2-5} 
                                           & $8^3$ $\times$ 8  & {[}0.5-0.8{]}          & 0.01  & $100\times 10^7$               \\ \hline
\multirow{3}{*}{0.3}                       & $4^3$ $\times$ 8  & {[}0.6-1.0{]}          & 0.01  & $100\times 10^7$               \\ \cline{2-5} 
                                           & $6^3$ $\times$ 8  & {[}0.6-1.0{]}          & 0.01  & $100\times 10^7$               \\ \cline{2-5} 
                                           & $8^3$ $\times$ 8  & {[}0.6-1.0{]}          & 0.01  & $100\times 10^7$               \\ \hline
\multirow{3}{*}{0.4}                       & $4^3$ $\times$ 8  & {[}0.6-1.0{]}          & 0.01  & $100\times 10^7$               \\ \cline{2-5} 
                                           & $6^3$ $\times$ 8  & {[}0.6-1.0{]}          & 0.01  & $100\times 10^7$               \\ \cline{2-5} 
                                           & $8^3$ $\times$ 8  & {[}0.6-1.0{]}          & 0.01  & $100\times 10^7$               \\ \hline
\multirow{3}{*}{0.5}                       & $4^3$ $\times$ 8  & {[}0.6-1.0{]}          & 0.01  & $100\times 10^7$               \\ \cline{2-5} 
                                           & $6^3$ $\times$ 8  & {[}0.6-1.0{]}          & 0.01  & $100\times 10^7$               \\ \cline{2-5} 
                                           & $8^3$ $\times$ 8  & {[}0.6-1.0{]}          & 0.01  & $100\times 10^7$               \\ \hline
\multirow{3}{*}{0.6}                       & $4^3$ $\times$ 8  & {[}0.6-0.8{]}          & 0.01  & $100\times 10^6$               \\ \cline{2-5} 
                                           & $6^3$ $\times$ 8  & {[}0.6-0.8{]}          & 0.01  & $100\times 10^6$               \\ \cline{2-5} 
                                           & $8^3$ $\times$ 8  & {[}0.6-0.8{]}          & 0.01  & $100\times 10^6$               \\ \hline
\multirow{3}{*}{0.7}                       & $4^3$ $\times$ 8  & {[}0.7-0.9{]}          & 0.01  & $100\times 10^6$               \\ \cline{2-5} 
                                           & $6^3$ $\times$ 8  & {[}0.7-0.9{]}          & 0.01  & $100\times 10^6$               \\ \cline{2-5} 
                                           & $8^3$ $\times$ 8  & {[}0.7-0.9{]}          & 0.01  & $100\times 10^6$               \\ \hline
\multirow{3}{*}{0.8}                       & $4^3$ $\times$ 8  & {[}0.8-0.9{]}          & 0.01  & $100\times 10^6$               \\ \cline{2-5} 
                                           & $6^3$ $\times$ 8  & {[}0.8-0.9{]}          & 0.01  & $100\times 10^6$               \\ \cline{2-5} 
                                           & $8^3$ $\times$ 8  & {[}0.8-0.9{]}          & 0.01  & $100\times 10^6$               \\ \hline
\multirow{3}{*}{0.9}                       & $4^3$ $\times$ 8  & {[}0.9-1.0{]}          & 0.01  & $100\times 10^6$               \\ \cline{2-5} 
                                           & $6^3$ $\times$ 8  & {[}0.9-1.0{]}          & 0.01  & $100\times 10^6$               \\ \cline{2-5} 
                                           & $8^3$ $\times$ 8  & {[}0.9-1.0{]}          & 0.01  & $100\times 10^6$               \\ \hline
\multirow{3}{*}{1.0}                       & $4^3$ $\times$ 8  & {[}0.9-1.1{]}          & 0.01  & $100\times 10^6$               \\ \cline{2-5} 
                                           & $6^3$ $\times$ 8  & {[}0.9-1.1{]}          & 0.01  & $100\times 10^6$               \\ \cline{2-5} 
                                           & $8^3$ $\times$ 8  & {[}0.9-1.1{]}          & 0.01  & $100\times 10^6$               \\ \hline
\multirow{3}{*}{1.1}                       & $4^3$ $\times$ 8  & {[}1.0-1.1{]}          & 0.01  & $100\times 10^6$               \\ \cline{2-5} 
                                           & $6^3$ $\times$ 8  & {[}1.0-1.1{]}          & 0.01  & $100\times 10^6$               \\ \cline{2-5} 
                                           & $8^3$ $\times$ 8  & {[}1.0-1.1{]}          & 0.01  & $100\times 10^6$               \\ \hline
\multirow{3}{*}{1.2}                       & $4^3$ $\times$ 8  & {[}1.1-1.2{]}          & 0.01  & $100\times 10^6$               \\ \cline{2-5} 
                                           & $6^3$ $\times$ 8  & {[}1.1-1.2{]}          & 0.01  & $100\times 10^6$               \\ \cline{2-5} 
                                           & $8^3$ $\times$ 8  & {[}1.1-1.2{]}          & 0.01  & $100\times 10^6$               \\ \hline
\multirow{3}{*}{1.3}                       & $4^3$ $\times$ 8  & {[}1.1-1.3{]}          & 0.01  & $100\times 10^6$               \\ \cline{2-5} 
                                           & $6^3$ $\times$ 8  & {[}1.1-1.3{]}          & 0.01  & $100\times 10^6$               \\ \cline{2-5} 
                                           & $8^3$ $\times$ 8  & {[}1.1-1.3{]}          & 0.01  & $100\times 10^6$               \\ \hline
\multirow{3}{*}{1.4}                       & $4^3$ $\times$ 8  & {[}1.2-1.3{]}          & 0.01  & $100\times 10^6$               \\ \cline{2-5} 
                                           & $6^3$ $\times$ 8  & {[}1.2-1.3{]}          & 0.01  & $100\times 10^6$               \\ \cline{2-5} 
                                           & $8^3$ $\times$ 8  & {[}1.2-1.3{]}          & 0.01  & $100\times 10^6$               \\ \hline
\multirow{5}{*}{1.5}                       & $4^3$ $\times$ 8  & {[}1.28-1.31{]}        & 0.001 & $100\times 10^6$               \\ \cline{2-5} 
                                           & $6^3$ $\times$ 8  & {[}1.28-1.31{]}        & 0.001 & $100\times 10^6$               \\ \cline{2-5} 
                                           & $8^3$ $\times$ 8  & {[}1.28-1.31{]}        & 0.001 & $100\times 10^6$               \\ \cline{2-5} 
                                           & $10^3$ $\times$ 8 & {[}1.28-1.31{]}        & 0.001 & $100\times 10^5$               \\ \cline{2-5} 
                                           & $12^3$ $\times$ 8 & {[}1.28-1.31{]}        & 0.001 & $100\times 10^5$               \\ \hline
\multirow{5}{*}{1.6}                       & $4^3$ $\times$ 8  & {[}1.33-1.36{]}        & 0.001 & $100\times 10^6$               \\ \cline{2-5} 
                                           & $6^3$ $\times$ 8  & {[}1.33-1.36{]}        & 0.001 & $100\times 10^6$               \\ \cline{2-5} 
                                           & $8^3$ $\times$ 8  & {[}1.33-1.36{]}        & 0.001 & $100\times 10^6$               \\ \cline{2-5} 
                                           & $10^3$ $\times$ 8 & {[}1.33-1.36{]}        & 0.001 & $100\times 10^5$               \\ \cline{2-5} 
                                           & $12^3$ $\times$ 8 & {[}1.33-1.36{]}        & 0.001 & $100\times 10^5$               \\ \hline
\multirow{5}{*}{1.7}                       & $4^3$ $\times$ 8  & {[}1.37-1.41{]}        & 0.001 & $100\times 10^6$               \\ \cline{2-5} 
                                           & $6^3$ $\times$ 8  & {[}1.37-1.41{]}        & 0.001 & $100\times 10^6$               \\ \cline{2-5} 
                                           & $8^3$ $\times$ 8  & {[}1.37-1.41{]}        & 0.001 & $100\times 10^6$               \\ \cline{2-5} 
                                           & $10^3$ $\times$ 8 & {[}1.37-1.41{]}        & 0.001 & $100\times 10^5$               \\ \cline{2-5} 
                                           & $12^3$ $\times$ 8 & {[}1.37-1.41{]}        & 0.001 & $100\times 10^5$               \\ \hline
\multirow{5}{*}{1.8}                       & $4^3$ $\times$ 8  & {[}1.42-1.45{]}        & 0.001 & $100\times 10^6$               \\ \cline{2-5} 
                                           & $6^3$ $\times$ 8  & {[}1.42-1.45{]}        & 0.001 & $100\times 10^6$               \\ \cline{2-5} 
                                           & $8^3$ $\times$ 8  & {[}1.42-1.45{]}        & 0.001 & $100\times 10^6$               \\ \cline{2-5} 
                                           & $10^3$ $\times$ 8 & {[}1.42-1.45{]}        & 0.001 & $100\times 10^5$               \\ \cline{2-5} 
                                           & $12^3$ $\times$ 8 & {[}1.42-1.45{]}        & 0.001 & $100\times 10^5$               \\ \hline
\multirow{5}{*}{1.9}                       & $4^3$ $\times$ 8  & {[}1.46-1.50{]}        & 0.001 & $100\times 10^6$               \\ \cline{2-5} 
                                           & $6^3$ $\times$ 8  & {[}1.46-1.50{]}        & 0.001 & $100\times 10^6$               \\ \cline{2-5} 
                                           & $8^3$ $\times$ 8  & {[}1.46-1.50{]}        & 0.001 & $100\times 10^6$               \\ \cline{2-5} 
                                           & $10^3$ $\times$ 8 & {[}1.46-1.50{]}        & 0.001 & $100\times 10^5$               \\ \cline{2-5} 
                                           & $12^3$ $\times$ 8 & {[}1.46-1.50{]}        & 0.001 & $100\times 10^5$               \\ \hline
\multirow{5}{*}{2.0}                       & $4^3$ $\times$ 8  & {[}1.50-1.53{]}        & 0.001 & $100\times 10^6$               \\ \cline{2-5} 
                                           & $6^3$ $\times$ 8  & {[}1.50-1.53{]}        & 0.001 & $100\times 10^6$               \\ \cline{2-5} 
                                           & $8^3$ $\times$ 8  & {[}1.50-1.53{]}        & 0.001 & $100\times 10^6$               \\ \cline{2-5} 
                                           & $10^3$ $\times$ 8 & {[}1.50-1.53{]}        & 0.001 & $100\times 10^5$               \\ \cline{2-5} 
                                           & $12^3$ $\times$ 8 & {[}1.50-1.53{]}        & 0.001 & $100\times 10^5$               \\ \hline
\multirow{3}{*}{2.5}                       & $4^3$ $\times$ 8  & {[}1.50-1.53{]}        & 0.001 & $100\times 10^6$               \\ \cline{2-5} 
                                           & $6^3$ $\times$ 8  & {[}1.50-1.53{]}        & 0.001 & $100\times 10^6$               \\ \cline{2-5} 
                                           & $8^3$ $\times$ 8  & {[}1.50-1.53{]}        & 0.001 & $100\times 10^6$               \\ \hline
\multirow{3}{*}{3.0}                       & $4^3$ $\times$ 8  & {[}1.50-1.53{]}        & 0.001 & $100\times 10^6$               \\ \cline{2-5} 
                                           & $6^3$ $\times$ 8  & {[}1.50-1.53{]}        & 0.001 & $100\times 10^6$               \\ \cline{2-5} 
                                           & $8^3$ $\times$ 8  & {[}1.50-1.53{]}        & 0.001 & $100\times 10^6$               \\ \hline
\multicolumn{1}{|l|}{\multirow{3}{*}{3.5}} & $4^3$ $\times$ 8  & {[}1.50-1.53{]}        & 0.001 & $100\times 10^6$               \\ \cline{2-5} 
\multicolumn{1}{|l|}{}                     & $6^3$ $\times$ 8  & {[}1.50-1.53{]}        & 0.001 & $100\times 10^6$               \\ \cline{2-5} 
\multicolumn{1}{|l|}{}                     & $8^3$ $\times$ 8  & {[}1.50-1.53{]}        & 0.001 & $100\times 10^6$               \\ \hline
\caption{
\vspace{3mm}
\label{SCTable}
Parameters for the Monte Carlo runs to determine the nuclear transition at strong coupling, with statistics after thermalization.}
\end{longtable}

\subsubsection{Finite $\beta$}
\label{OBTable}
All runs at finite $\beta$ have been obtained for $\Nt=4$, which corresponds to a moderately low temperature $aT=0.25$ compared to the value of the chiral transition $aT\simeq 1.54$. Those simulations were too expensive to attempt $\Nt=8$ runs, in particular as a higher statistics was required.

The spatial volumes are $4^3$, $6^3$ and $8^3$. For $\beta$ values are from $0.0$ to $1.0$ with step size $0.1$, and for $am_q$ values from $0.00$ to $1.00$ with step size $0.01$. The values of $a\mu$ were chosen close to the nuclear transition, the scanning range is shifted to large values as $am_q$ increases. At small quark masses the scanning range is from $a\mu=0.4$ to $1.0$ and for the large quark masses, it is from $0.6$ to $1.2$ with step size $0.01$.
The statistics used for are $15\times 10^4$ measurements and between measurement, $40 \times N_s^3$ worm updates.  

\bibliography{nucleartransition} 
\bibliographystyle{apsrev4-1}

\end{document}